\documentclass[review,12pt]{elsarticle}

\usepackage{lineno,hyperref}
\modulolinenumbers[5]

\usepackage{hyperref}

\usepackage{color}
\definecolor{darkblue}{rgb}{0.,0.,0.4}
\definecolor{darkred}{rgb}{0.5,0.,0.}

\journal{arXiv}

\usepackage{url}  
\usepackage{graphicx} 
\usepackage{revsymb}
\usepackage{bm}
\usepackage{textcomp}
\usepackage{revsymb}
\usepackage{amssymb}
\usepackage{ifsym}
\usepackage{wasysym}
\usepackage{txfonts}
\usepackage{hyperref}
\usepackage{tipa}
\usepackage{cleveref}[2012/02/15]   
\crefformat{footnote}{#2\footnotemark[#1]#3}
\usepackage{multirow}
\usepackage{caption}

\begin{document}
	
	\begin{frontmatter}
		
		\title{%
			Thermal-fluctuation mechanism of long-term corrosion of glass caused by internal stress: processes of depolymerization, impurity migration and fracturing on an atomic scale
		}

		
		

		
		\author[GOSNIIR]{Irina~F.~Kadikova}
		\ead{kadikovaif@gosniir.ru}
		
		\author[GOSNIIR]{Tatyana~V.~Yuryeva} 
		\ead{yuryevatv@gosniir.ru}

		\author[GOSNIIR]{Ekaterina~A.~Morozova}
		\ead{morozovaea@gosniir.ru}
		
		\author[Hermitage]{Irina~A.~Grigorieva}
		\ead{grigorieva\_ia@hermitage.ru}

		\author[RF_CFS]{Ilya~B.~Afanasyev}
		\ead{il.afanasyev@sudexpert.ru}
		
		\author[MINERAL]{Vladimir~Y.~Karpenko}
		\ead{mineralab@mail.ru}
		
		\author[GPI]{Vladimir~A.~Yuryev\corref{correspondingauthor}}
		\ead{vyuryev@kapella.gpi.ru}
		
		\date{}	
		
		\address[GOSNIIR]{The State Research Institute for Restoration, Building~1, 44~Gastello Street, Moscow 107114, Russia}
		
		
		\address[Hermitage]{The State Hermitage Museum, 34~Dvortsovaya Embankment, Saint Petersburg 190000, Russia}
		
		
		
		
		
		\address[RF_CFS]{The Russian Federal Center of Forensic Science of the Ministry of Justice, Building~2, 13~Khokhlovskiy Side~Street, Moscow 109028, Russia}
		
		
		\address[MINERAL]{A.\,E.\,Fersman Mineralogical Museum of the Russian Academy of Sciences, Building~2, 18~Leninsky Avenue, Moscow 119071, Russia}
		
		\address[GPI]{A.\,M.\,Prokhorov General Physics Institute of the Russian Academy of Sciences, 38~Vavilov Street, Moscow 119991, Russia}

		\cortext[correspondingauthor]{Corresponding author}
		%



\begin{abstract} 
The long-term corrosion of glass is thoroughly investigated using polarising light microscopy, FTIR spectroscopy and scanning electron microscopy including EDX spectroscopy and elemental mapping.
Early 19th century beads made of turquoise lead-potassium glass are the object of the study.
Considerable non-uniformly distributed internal stresses were introduced into the studied glass during bead manufacturing.
The corrosion of turquoise seed-bead glass of the 19th century has been shown to comprise several mutually connected processes developing in parallel and intensifying one another.
These processes are the depolymerization of glass, the directed migration of alkali-metal impurities and local glass leaching, and the nucleation of micro discontinuities followed by the formation of micro cracks and their gradual growth.
In course of time, their cumulative and apparently synergistic effect results in glass fracturing terminated with bead crumbling into tiny discoloured sand-like particles.
All these phenomena are controlled by the common driving force, namely, by the internal stress originated from the bead production process.
Thus, the phenomenon of corrosion of beads made of stressed glass is the process of the internal stress relaxation in glass through the nucleation and growth of cracks, which lasts for many decades at room temperature.
It is dramatically accelerated, however, when beads are heat treated at moderate temperature.
Annealing at {300\textcelsius} for 15~minutes appeared to be sufficient for artificial ageing of initially intact beads.
The corrosion has been observed to occur in glass even if beads are kept in well-controlled museum environment.
The local thermal-fluctuation mechanism has been proposed to account for the corrosion phenomenon on an atomic scale.
The approach based on the thermal-fluctuation mechanism is applicable to describe degradation and destruction processes in various solid substances and composite materials under internal or external stress; 
this approach enables understanding the phenomena of formation and degradation of nanostructures, in particular, heterostructures with quantum wells or quantum dots.
\end{abstract}

\begin{keyword}
	\texttt{Thermal-fluctuation mechanism \sep Stress corrosion\sep Glass corrosion \sep Glass depolymerization \sep Impurity diffusion \sep Impurity segregation \sep Glass cracking }
\end{keyword}

\newpage

\begin{graphicalabstract}\\
	\includegraphics[width=\textwidth]{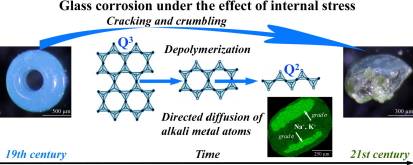}
\end{graphicalabstract}

\begin{highlights}
	\item Intact and degraded domains are analysed in the same glass beads using FTIR and SEM
	\item Glass depolymerizes in degraded domains and does not depolymerize in intact ones
	\item Alkali elements migrate from degraded domains to intact ones due to internal stress
	\item Internal stresses cause glass depolymerization, bleaching, cracking and fragmenting
	\item Atomic-level thermal-fluctuation mechanism of glass stress corrosion is discussed
\end{highlights}

\end{frontmatter}


\newpage

\section{Introduction} \label{sec:intro}

Glass resistance to chemical agents, such as water, acids or alkalies, has been a subject of extensive investigations in glass science and technology since the time of the very beginning of technical glass manufacturing  \cite{Glasses_durability,Chemical_Durability_Optical_Glass,Glass_Durability_thermodynamic,Molecular_mechanisms_corrosion}.
Presently, this issue, as well as techniques for testing and inspection of glass chemical stability, are usually considered in monographs and textbooks on glass manufacturing technology (see, e.g., Refs.~\cite{Glass_chemical_resistance,GLASS_Analysis_Corrosion}).
The problem of glass deterioration under the effect of aggressive environment, primarily of atmospheric moisture, is also highly important in science and practice of the cultural heritage conservation \cite{Degradation_Glass,Glass_corrosion_multiscale,Robinet_1,Robinet_2,Robinet_3,Robinet_4,Unstable_Historic_Glass,Corrosion_Sci,Long-term_modeling_alteration-transport}.
The process of surface chemical corrosion of artistic or historical glasses, which is caused mainly by the effect of air moisture or acidic ambient during keeping resulting in surface layer leaching, is often referred to as `crizzling' \cite{Crizzling_Problem_1975,Crizzling_Problems_Solutions}. 
Crizzling is frequently considered as the only corrosion mechanism for all kinds of historical glass including bead glass \cite{Postprints,BEAD_FORUM,Etn_Conc_Newsletter,Book_Adam_Lovell,Canadian_beads_Deterioration}.
Craquelure that sometimes appears on surfaces of beads made of some kinds of glass, mainly of potassium glass, due to long-term keeping or exploitation is considered as the outcome of crizzling, i.e. glass leaching at the surface. 

Some time ago, during the restoration of beaded articles made from early to mid 19th century, kept in collections of Russian museums, we found that tiny semi-opaque turquoise seed beads%
\footnote{%
	They are sometimes also referred to as blue-green beads due to their tendency to turn greener as their glass degrades.%
}%
%
---nearly 1 mm in diameter---%
were the most susceptible to heavy wear and tear, with both intact ones and those at various phases of degradation usually being in the immediate neighbourhood in an embroidery (see Online Resource~\ref{esm:movie}) \cite{Yuryev_JOPT}. 
Moreover, adjacent beads in a historical beaded article were often seen to be at radically different phases of corrosion---heavily degraded beads often adjoined intact ones while they were obviously very similar and had close chemical composition \cite{Yur_JAP}.
In addition, in our previous works, we succeeded to find out that bead breaking always begins in the core rather than on the surface and some beads looking intact contain cracks in their cores (we refer to this phenomenon as `internal corrosion') \cite{Yur_JAP,Beads_KSS_SPIE,KSS_Electron_microscopy}.
These facts made us suppose that some physical processes other than crizzling might be responsible for the corrosion of seed beads since analogous adjacent beads must be affected by chemical agents to the equal degree and, after nearly one and a half century of utilization of the beaded articles in everyday life followed by keeping in a museum, all the beads must be corroded to the close extent.%
\footnote{%
Besides, we failed to detect any products of chemical corrosion on the surface of beads (e.g., alkali in the form of white crystals or droplets of liquid alkali) \cite{2020_Glas_depolimer}, which would be present if surface chemical reactions characteristic of crizzling took place.
}

Comprehensive investigations of the long-term corrosion of 19th century semi-opaque turquoise seed beads have elucidated the mechanism of this phenomenon \cite{Yur_JAP,Beads_KSS_SPIE,KSS_Electron_microscopy,2020_Glas_depolimer}.
We have concluded that nothing but local stresses in glass are responsible for the long-term corrosion of this kind of seed beads \cite{Beads_KSS_SPIE,KSS_Electron_microscopy} (see Online Resource~\ref{esm:movie}).
There are two sources of the local stress in the turquoise bead glass.
The first one is orthorhombic KSbOSiO$_4$ (KSS) micro crystallites and their colonies (Online Resource~\ref{esm:fig_KPbSbO5}, Fig.~ESM~\ref{esm:fig_KPbSbO5}.1) that arose in glass during its melting and bubbling from K$^+$ and Sb$^{5+}$, which were present in the glass melt in high enough concentrations \cite{KSS_Electron_microscopy}.
Because of the difference in temperature coefficients of linear expansion of glass and KSS crystals, the colonies of KSS crystals generated tensile strain in the glass matrix during bead cooling down after tumble finishing giving rise to micro discontinuities in the internal domains (cores) of the beads \cite{KSS_Electron_microscopy}.
In course of time, the discontinuities slowly merged and elongated, transformed to micro cracks and then to long fractures (Online Resource~\ref{esm:fig_KPbSbO5}, Fig.~ESM~\ref{esm:fig_KPbSbO5}.1).
Eventually beads crumbled to corroded particles and fallen out of beaded embroideries.
At room temperature, this process might last for decades and centuries.

At the same technological step of bead production, shells of re-vitrified glass arose at surfaces of beads (Online Resource~\ref{esm:fig_KPbSbO5}, Fig.~ESM~\ref{esm:fig_KPbSbO5}.1\,a,\,b) \cite{KSS_Electron_microscopy}.
Glass softening near surface during tumble polishing at elevated temperature followed by its repeated vitrification that has been detected due to imprints of polishing powder particles at the walls of bead holes gave rise to these domains \cite{KSS_Electron_microscopy}.
The thickest shell regions were formed around bead holes since they were filled with a mixture of clay, chalk and charcoal during tumbling and cooling \cite{Yurova_large} and hence glass was heat insulated from the ambient at holes therefore warmed up more and cooled down slower than that at the rest parts of beads.
After cooling and repeated vitrification, these domains were formed by denser glass than the cores \cite{Glass_Technology_Ch.7}.
Outer shells were thinner than those around holes and likely quenched due to better heat exchange with the ambient.
Cores and shells obviously should differ in temperature coefficients of linear expansion and this gave rise to the additional internal stress at the core-shell interface and in the shells.
That is the second source of the internal stress in bead glass.
Eventually, fractures arose at the core-shell interface (Online Resource~\ref{esm:fig_KPbSbO5}, Fig.~ESM~\ref{esm:fig_KPbSbO5}.1\,a,\,b).
The formation of micro cracks in glass shells, especially around bead holes, was also caused by stress originating from bead cooling after tumbling.
Further growth of cracks, which eventually resulted in bead fragmentation, went on due to the gradual relaxation of the residual stress in glass \cite{KSS_Electron_microscopy}.

The model of the internal-stress-induced glass disease explains the presence of adjacent beads at different phases of corrosion in beaded articles.
If these beads were cooled at different rates after tumble finishing, internal stresses of different magnitude emerged in them, especially around orifices.
Therefore, they degraded at different rates, and the greater the bead cooling rate was, the higher the glass degradation rate was.
Besides that, the internal-stress-driven glass degradation model elucidates the long-term internal corrosion by the effect of KSS colonies.  

Thus, now we consider gradual, for the period of decades, fracturing of seed-bead glass as a kinetic process of consecutive release of the internal stress rather than an outcome of chemical reactions on the surface \cite{KSS_Electron_microscopy}.

We have also reported another phenomenon related to the decay of seed-bead glass due to the internal stress \cite{2020_Glas_depolimer,RCEM-2018_vibrational_spectroscopy,Preservation_Cultural_Heritage-2018,GosNIIR_Conference}.
Investigating semi-opaque turquoise seed beads at different stages of corrosion by the Fourier-transform infrared spectroscopy we have found out that more corroded beads demonstrate more depolymerized structure of glass. 
Vibrations assigned to SiO$_4$ chains (Q$^2$) and paired SiO$_4$ tetrahedra (Q$^1$) have been demonstrated to predominate in the glass structure of strongly corroded beads whereas a wide set of vibrational bands with close intensities and more uniform distribution of spectral components assigned to different silicate structures has been detected in well-preserved beads.
The degradation of the glass structure may reduce its strength \cite{Structure_strength_glasses} and promote the nucleation of micro cracks in its volume as well as the growth of existing ones.
We have also concluded \cite{2020_Glas_depolimer} that the scenario attributing bead-glass internal corrosion to local strains, which had previously been proposed by us to explain the long-term degradation process of turquoise bead glass at room temperature \cite{KSS_Electron_microscopy}, has agreed with the commonly adopted thermal fluctuation theory of materials fracture \cite{Atomic-Level_Fracture_Solids,Beads_EAMC-2018}. 
The simplification of the glass structure under internal stress during the long-term degradation of glass at room temperature also finds a realistic explanation within this theory. 
We have suggested that glass depolymerization, caused by the internal stress and decreasing the glass strength, is an essential corrosion mechanism of stressed glass \cite{2020_Glas_depolimer}.

In the present work, we have carried out the comprehensive investigation of the glass depolymerization phenomenon to verify the previously proposed hypothesis employing more reliable and accurate experimental approach based on a much broader statistics.
We have been convinced using polarizing microscopy that considerable non-uniformly distributed internal stresses had been introduced into the studied glass by the process of bead manufacturing.
We have compared infrared absorption spectra recorded at numerous domains of different degree of degradation---from intact to heavily damaged ones---in the same samples of beads, each of which, in turn, had also been subjected to glass corrosion to different extent.
We unambiguously demonstrate that glass is depolymerized in degraded domains of beads, while it remains its polymerized structure in intact ones even if both degraded and intact domains are in the same samples and, moreover, are adjacent.     
Additionally, we have investigated maps of distribution of chemical elements in these beads acquired by means of the X-ray energy dispersive micro spectroscopy to compare the distribution of chemical elements, primarily alkali metals, silicon and oxygen, in intact and degraded domains.
We demonstrate that alkali metals leave degraded domains moving into intact ones, while the Si--O matrix becomes denser in the degraded regions. 
Moreover, we have discovered that alkali metals migrate along the stress gradient and accumulate at domains with the maximum stress, whereas the silicate matrix has a density minimum in the same regions.  
Finally, we demonstrate the accelerated degradation of beads as a result of heat treatments at elevated temperature---annealing at {300\textcelsius} for 15~minutes appeared to be sufficient for \textit{`artificial ageing'} of initially intact beads---as well as the effect of sample polishing on the rate of glass deterioration, which evidence in support of the proposed model of corrosion induced by the internal stress.

We discuss the experimental observations and come to the conclusion that the local thermal-fluctuation mechanism \cite{2020_Glas_depolimer,Atomic-Level_Fracture_Solids} explains on an atomic scale all the discovered processes that together constitute the phenomenon of stressed-glass corrosion.

We should emphasize that the approach based on the thermal-fluctuation mechanism enables the physical description of degradation and destruction processes in various solid substances and composite materials under the action of internal or external stress. 
This approach allows one to understand the phenomena of formation and degradation of nanostructures, in particular, heterostructures with quantum wells or quantum dots \cite{GE_QD-strain_above,Peculiarities_Raman}.

Now, after the extended preface, we proceed to the presentation of the results of our new experiments.

\section{Sample preparation, experimental methods and equipment} \label{sec:experiment}

\subsection{Samples} \label{subsec:samples}

\subsubsection{Samples of beads}  \label{subsec:bead_samples} 

Samples of glass seed beads were obtained from beaded articles of the 19th century. 
All samples were washed with high purity isopropyl alcohol ([C$_3$H$_7$OH]~$>99.8$~wt.\,\%) at 40{\textcelsius} for 20 minutes in a chemical glass placed into an ultrasonic bath filled with water ($\nu=40$~kHz, $P=120$~W) \cite{Beads_KSS_SPIE,2020_Glas_depolimer}. 

The analyses were mainly carried out using sections of bead samples, 
which were prepared following a standard procedure.
First, clean bead samples were immersed in epoxy resin L;%
\footnote{%
	For the IR absorbance spectrum of epoxy resin L, see Ref.~\cite{2022_Glas_alter_ESM-2}.
}
then, after hardening, the preparations were polished using corundum paste for grinding and diamond paste for finishing until thin smooth plates were obtained.
After polishing, the sections were washed in isopropyl alcohol.%

Some samples of beads were annealed for artificial ageing at {300\textcelsius} for 15~minutes in the atmospheric air in a muffle furnace.
We studied these samples without the preparation of thin sections. 

Glass sections were not made from bead samples used for the internal stress investigation either.

\subsubsection{Synthesis and analysis of KSbOSiO$_4$ reference samples} \label{sec:KSS_synthesis}

Reference samples of the KSbOSiO$_4$ were synthesized following the process described in Ref.\,\cite{COD-KSbOSiO4}: 
KNO$_3$, Sb$_2$O$_3$ and SiO$_2$ taken in stoichiometric proportion were annealed at the temperature of 1100{\textcelsius} in air ambient for 24 hours. 
The resultant samples were powder of polycrystalline particles of various sizes (Online Resource~\ref{esm:fig_KPbSbO5}, Fig.~ESM~\ref{esm:fig_KPbSbO5}.2).

X-ray powder diffraction pattern of the obtained compound verified that a single phase of orthorhombic KSbOSiO$_4$ was synthesized \cite{Beads_KSS_SPIE,KSS_Electron_microscopy}. 
Peaks related to the tetragonal phase \cite{COD-KSbOSiO4,Tetragonal-orthorhombic} were not observed in the powder patterns as well as peaks, which could be attributed to other substances. 

Raman spectra of the synthesized standard of orthorhombic KSbOSiO$_4$ are presented in Refs.\,\cite{Beads_KSS_SPIE,Raman_KSS_Brandt,KSS_Raman_spectrum}. 
Their characteristic feature is an intense narrow peak at about 532~cm$^{-1}$.
Cathodoluminescence spectra of orthorhombic KSbOSiO$_4$ acquired at room and liquid nitrogen temperatures are demonstrated in Ref.\,\cite{KSS_Electron_microscopy}; the characteristic band of KSS has two maxima at about 444 and
483 nm.

We present FTIR spectra and absorption bands of orthorhombic KSbOSiO$_4$ in Online Resource~\ref{esm:fig_KPbSbO5} (Figs.~ESM~\ref{esm:fig_KPbSbO5}.3 to ESM~\ref{esm:fig_KPbSbO5}.5, Tables~ESM~\ref{esm:fig_KPbSbO5}.1 and ESM~\ref{esm:fig_KPbSbO5}.2).

\subsection{Methods and instruments} \label{subsec:methods}

Fourier transform infrared (FTIR) spectral analysis of bead samples and KSS in the spectral range from 600 to 4000~cm$^{-1}$  was performed using a LUMOS microscope (Bruker) in the attenuated total reflection (ATR) mode with a Ge ATR crystal.
FTIR spectra of KSS recorded in the interval between 50 and 4000~cm$^{-1}$ were obtained using a Vertex 70v (Bruker) spectrometer in the ATR mode with a diamond ATR crystal.  
Experiments were carried out at the spectral resolution of 4~cm$^{-1}$; routinely, 64 scans were averaged for each spectrum. 

Scanning electron microscopy (SEM) images were usually obtained using Vega-II~XMU, Mira~3~LMH or Mira~3~XMU microscopes (Tescan Orsay Holding). 
An image of the KSS powder was obtained using JSM-7001F SEM (Jeol).
As a rule, SEM images were recorded both in backscattered electrons (BSE) and secondary electrons (SE) since SEM operated in the former mode represents mainly the substance elemental composition (i.e. its density), while operated in the latter mode it renders mainly the spatial relief, whereas both types of the information are usually required to correctly interpret data obtained using SEM.

Elemental microanalysis and mapping were made with Inca Energy~450, EDS X-MAX~50 or AZtecOne energy dispersive X-ray (EDX) spectrometers (Oxford Instruments).
Some analyses were carried out using TM4000 Plus SEM (Hitachi) equipped with Quantax 75 (Bruker) EDX spectrometer.

For the examination of bead samples using light microscopy (LM), including polarizing microscopy (PM), POLAM L213-M (LOMO) polarizing microscopes were employed.

An Ortholux 2 POL (Leitz Wetzlar) polarizing microscope was used to study samples in the immersion liquid; 
during the experiments, the immersion liquids were being selected from a set with the refractive index $n$ from 1.400 to 1.700 in such a way to equalize the sample glass and liquid refractive indices for minimizing the light internal reflection in the samples under the study.  
To compare the immersion liquid refractive index with that of the studied sample, the Becke method was used (see, e.g., \cite{Becke_line_method-en,Becke_line_test-ru,Becke_line_method-geol}). 
Immersion liquids had been being changed in the cuvette until a bright rim at the interface of the sample and liquid visually disappeared that was considered as an evidence for the equality of their refractive indices.

MSP-1 transmission light microscopes (LOMO) were also used for micro photographing of the samples.

A SNOL-8,2/1100 laboratory muffle furnace (AB UMEGA-GROUP) was used for bead thermal treatments.

\section{Experimental results}    \label{sec:results}

\subsection{Stress in beads}  \label{subsec:stress}  

\subsubsection{Test with a needle} \label{subsubsec:needle} 

As mentioned above, the studied unstable translucent turquoise glass seed beads made in the 19th century mainly suffer high internal stress, with the magnitude of the stress determining the current preservation degree of beads by controlling the mean rate of glass degradation. 
The movie presented in Online~Resource~\ref{esm:movie} brightly demonstrates it.
Degraded beads are seen to instantly break, with glass pieces flying away like bullets, when touched with a needle, whereas similar intact ones, adjacent in a beadwork, do not even if intensely touched with a needle.
This effect, which resembles explosive disintegration of Prince Rupert's drops \cite{PRD_Royal_Soc_London-1921,PRD_Royal_Soc_London-1986,PRD_Phylos_Mag_B-1994,PRD_Phylos_Mag_Lett-1998,PRD_Appl_Phys_Lett-2018}, is due to a high internal stress accumulated in the deteriorated beads during making \cite{KSS_Electron_microscopy} as well as low strength of the fractured glass. 
The intact beads are not as stressed and for this reason do not contain internal cracks, and therefore they are much stronger than the deteriorated ones.

\subsubsection{Polarizing microscopy} \label{subsubsec:PM} 

\begin{figure}[th] 	
	\begin{minipage}[l]{\textwidth}
		\includegraphics[scale=1]{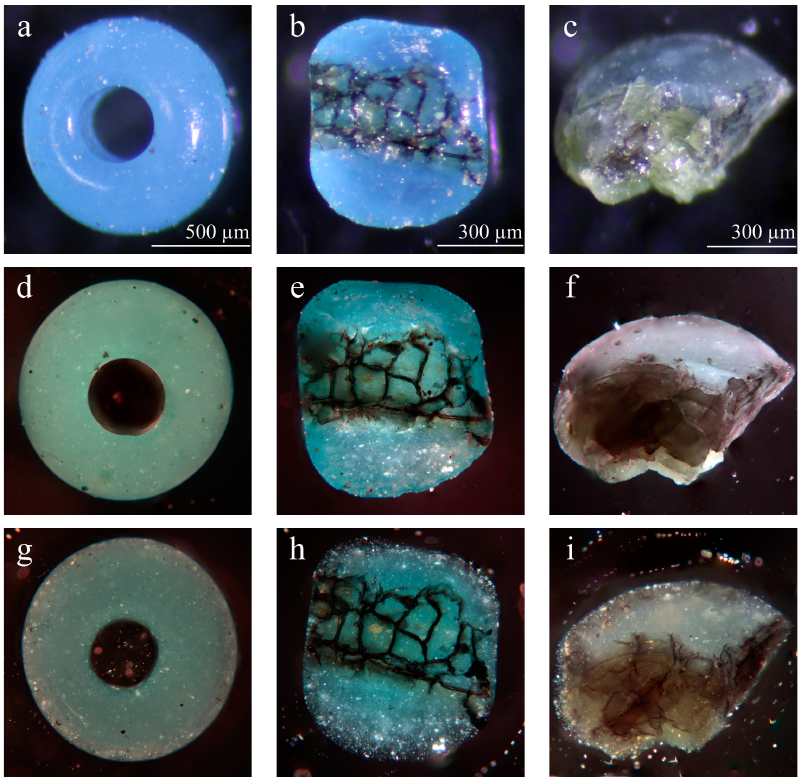}
	\end{minipage}
	\caption{(Colour online)
		LM images of three semi-opaque turquoise glass beads at different stages of corrosion  \cite{Yuryev_JOPT,Yur_JAP,Beads_KSS_SPIE}: 
		(a\,--\,c) 
		survey images with scale bars of 
		(a) an intact bead,
		(b) a piece of a moderately damaged bead fallen apart due to large cracks 
		and 
		(c) a particle of a heavily damaged (discoloured and fully crumbled) bead;
		PM images:
		(d\,--\,f)   
		crossed polars, no immersion liquid;
		(g\,--\,i) 
		crossed polars, immersion liquid, equalized refractive indices ($n=1.499$).
	}
	\label{fig:Beads_stress}	   
\end{figure}

Experiments using immersion polarizing microscopy allowed us to reveal internal stress in the studied glass beads (Fig.~\ref{fig:Beads_stress}). 
Here, we present as an example the results obtained for three bead samples at different phases of glass corrosion \cite{Yuryev_JOPT,Yur_JAP,Beads_KSS_SPIE} shown in Fig.~\ref{fig:Beads_stress}a\,--\,c: an undamaged bead, a moderately damaged one and a heavily damaged one.
PM photographs taken at crossed nicols in air without immersion in liquid demonstrate bright images with sharp edges
(Fig.~\ref{fig:Beads_stress}d\,--\,f), which are in fact dark-field images.
The rotated light polarization component, normal to the illuminating light polarization plane, which passes through the analyser to form the image, mainly arises due to light internal reflections from the sample surface.
Nevertheless, a part of light in the images may be associated with birefringence emerging due to the internal stress if the latter is present.
Using immersion in liquid and equalizing refractive indices of glass and liquid one can remove the internal reflection and extract light that passes through crossed nicols due to the internal stress.
This is easy to do for transparent glass but even for for semi-transparent one, the Becke method can be applied to determine the value of glass refractive index by comparing it with that of the chosen immersion liquid \cite{Becke_line_method-en}.

The Becke test has shown that the refractive indices of the studied samples were in the range $1.490<n<1.508$.
The Becke line at the glass-liquid interface visually completely disappeared in the immersion liquid with $n=1.499$ that allowed us to conclude that $n\approx 1.499$ for glass of all the samples in this experiment. 

Micro photographs of the samples immersed in the liquid with $n=1.499$ obtained at crossed polars are given in Fig.~\ref{fig:Beads_stress}g\,--\,i.
Thin edges of the bead pieces shown in the panels (h) and (i) are invisible;
only colonies of KSS crystals and other inhomogeneities, e.g., bubbles are seen brightly illuminated with light polarized normally to the polarization plane of the microscope illumination light.
Even the intact sample is well visible through the crossed nicols demonstrating numerous internal inclusions despite it looks turbid without the immersion.
This makes us conclude that we observe the internal stress that is responsible for birefringence in glass.

The stress is visible in the whole intact sample and in the whole particle of the moderately damaged sample.
Yet, in the particle of the crumbled bead, the stress is observed only in the relatively well preserved glass grains.
Small grains seen bright in the survey image (panel c) look dark in the polarized light without the immersion (panel f) and slightly illuminated when the sample is immersed in liquid (panel i).
This allows us to state that heavily cracked domains are much more relaxed compared to well-preserved ones.

\subsection{SEM SE and BSE imaging and light microscopy}   \label{subsec:EEM_SE-BSE}

\begin{figure}[th] 	
	\begin{minipage}[l]{\textwidth}{\hspace{-2.5cm}} 
		\includegraphics[scale=1]{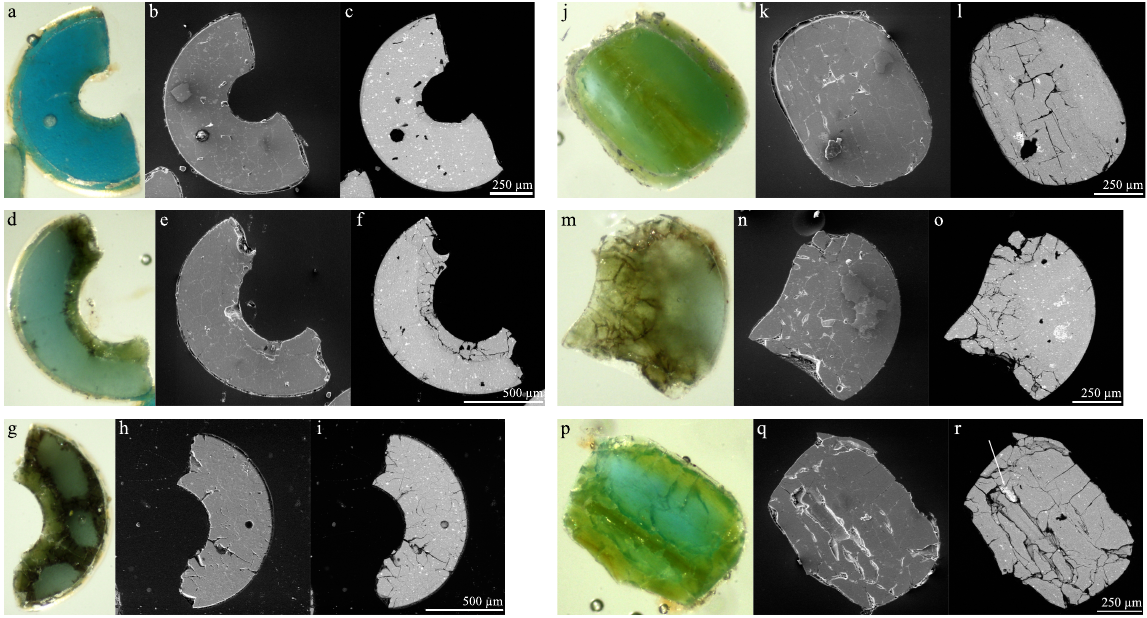}
	\end{minipage}
	\caption{(Colour online)
		Light and SEM micro images of the studied sections of semi-opaque turquoise glass beads:
		(a\,--\,c)~sample~1, 
		(d\,--\,f)~sample~2, 
		(g\,--\,i)~sample~3, 
		(j\,--\,l)~sample~4, 
		(m\,--\,o)~sample~5, 
		(p\,--\,r)~sample~6; 
		(a, d, g, j, m, p) transmission light,
		and SEM 
		(b, e, h, k, n, q) SE
		and
		(c, f, i, l, o, r) BSE images. 
	}
	\label{fig:Beads_samples}	   
\end{figure}

Fig.~\ref{fig:Beads_samples} demonstrates SEM images of the studied sections of semi-opaque tur\-quoise glass beads;
light images are also shown for readers to be able to compare them with the SEM ones.
Glass domains of different degree of destruction are seen in the images.
The sample 1 is seen to be practically undamaged throughout the whole volume except for a thin region surrounding its hole (panels a to c), which looks slightly greenish in the transmitting light image and is seen to contain numerous fine cracks in the SEM ones.
The next sample (panels d to f) is more corroded.
The region around its hole is thicker and much more fractured.
Besides, a domain of strongly degraded glass is observed at its upper end.
The sample 3 contains domains of both strongly and moderately damaged glass (panels g to i), with a network of fine cracks being observed on the surfaces of the latter ones.
Notice that three domains of blueish green glass that are seen in the light image are likely buried in the glass bulk a few micrometers under the section surface, which is more damaged that gives the glass a greener tint in transmitted light rays.
The sample 4 (panels j to l) is strongly corroded; yet, undamaged glass domains are seen to have remained in its interior: they are well seen on the left and on the right of the network of large cracks in the centre of its core.
A thick shell full of cracks frames the bead.
Finally, the samples 5 and 6 (panels m to o and p to q, respectively) are strongly degraded beads at the final phase of corrosion \cite{Yuryev_JOPT,Yur_JAP,Beads_KSS_SPIE}. 
They does not contain undamaged glass.
They are strongly cracked and discoloured.

We should note that bright white spots and groups of such spots in the BSE images represent crystals of orthorhombic KSS and their accumulations \cite{Yur_JAP,Beads_KSS_SPIE,KSS_Electron_microscopy}.
Large colonies of KSS are seen in Fig.~\ref{fig:Beads_samples}\,l,\,o and~r (one of the colonies is shown with an arrow in the panel~r).
More images of KSS crystals that rapture glass of turquoise beads are presented in
Online Resource~\ref{esm:fig_KPbSbO5}, Fig.~ESM~\ref{esm:fig_KPbSbO5}.1.
Core and shell domains of beads are also indicated in
Fig.~ESM~\ref{esm:fig_KPbSbO5}.1.

There is an important peculiarity of the spatial distribution of the bead glass composition that can be observed only by means of SEM operated at the BSE mode.
As is known, the magnitude of the BSE signal strongly depends on the density of the studied substance:
the higher is the substance density in some area, the more intensely the area reflects electrons and thus the brighter is its image.
Undamaged or slightly damaged glass domains are seen to look coloured somewhat paler grey than strongly degraded ones in the SEM BSE images (Fig.~\ref{fig:Beads_samples}\,c,\,f,\,l).
This means that they consist of heavier glass than strongly corroded domains do.
The reason of the observed inhomogeneity in the distribution of the glass density in the beads will be discussed below.

\begin{figure}[th] {\vspace{-4cm}}
	\begin{minipage}[l]{1.2\textwidth}{\hspace{-2cm}} 
		\includegraphics[scale=1]{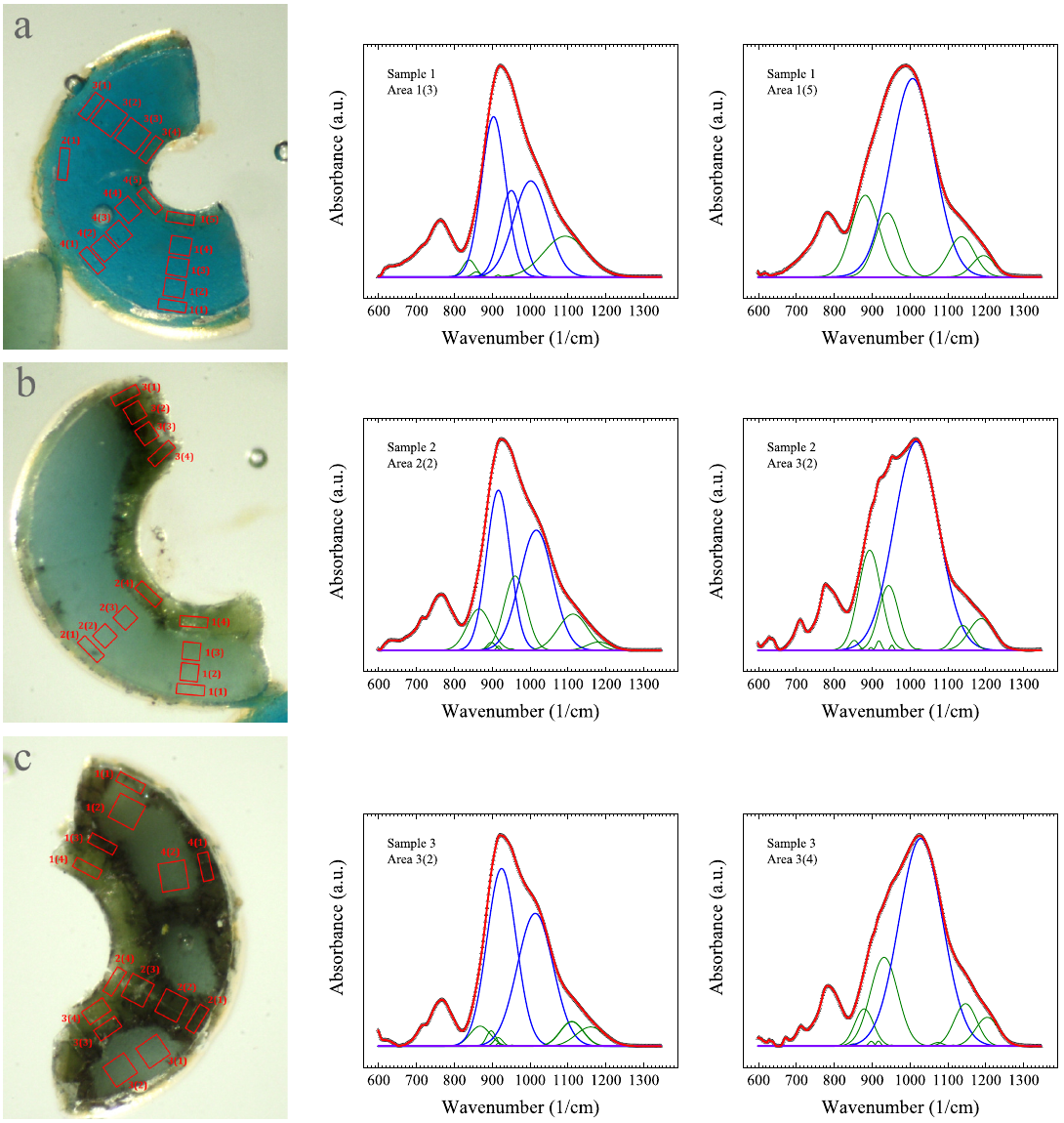}
		\caption{(Colour online) 
			Micro photographs and characteristic IR absorption spectra of the studied samples:
			(a)~sample~1, 
			(b)~sample~2, 
			(c)~sample~3, 
			(d)~sample~4, 
			(e)~sample~5  
			and
			(f)~sample~6; 
			numbered rectangles in the images show the areas of FTIR analysis;
			empty triangles in the plots depict the experimental data;
			olive and blue lines present peaks derived as a result of the spectra deconvolution, 
			with the blue ones showing the strongest of them,
			red lines are the cumulative fit curves,
			violet ones are baselines;
			moderate and weak peaks at the wave numbers below 840~cm$^{-1}$ are omitted;
			FTIR spectra for all the marked areas are given in Online Resource~\ref{esm:fig_FTIR_All_Spectra}.
		}
		\label{fig:Beads_FTIR}	       
\end{minipage}\end{figure}

\begin{figure}[th] {\vspace{-4cm}}
	\begin{minipage}[l]{1.2\textwidth}{\hspace{-2cm}}
			\includegraphics[scale=1]{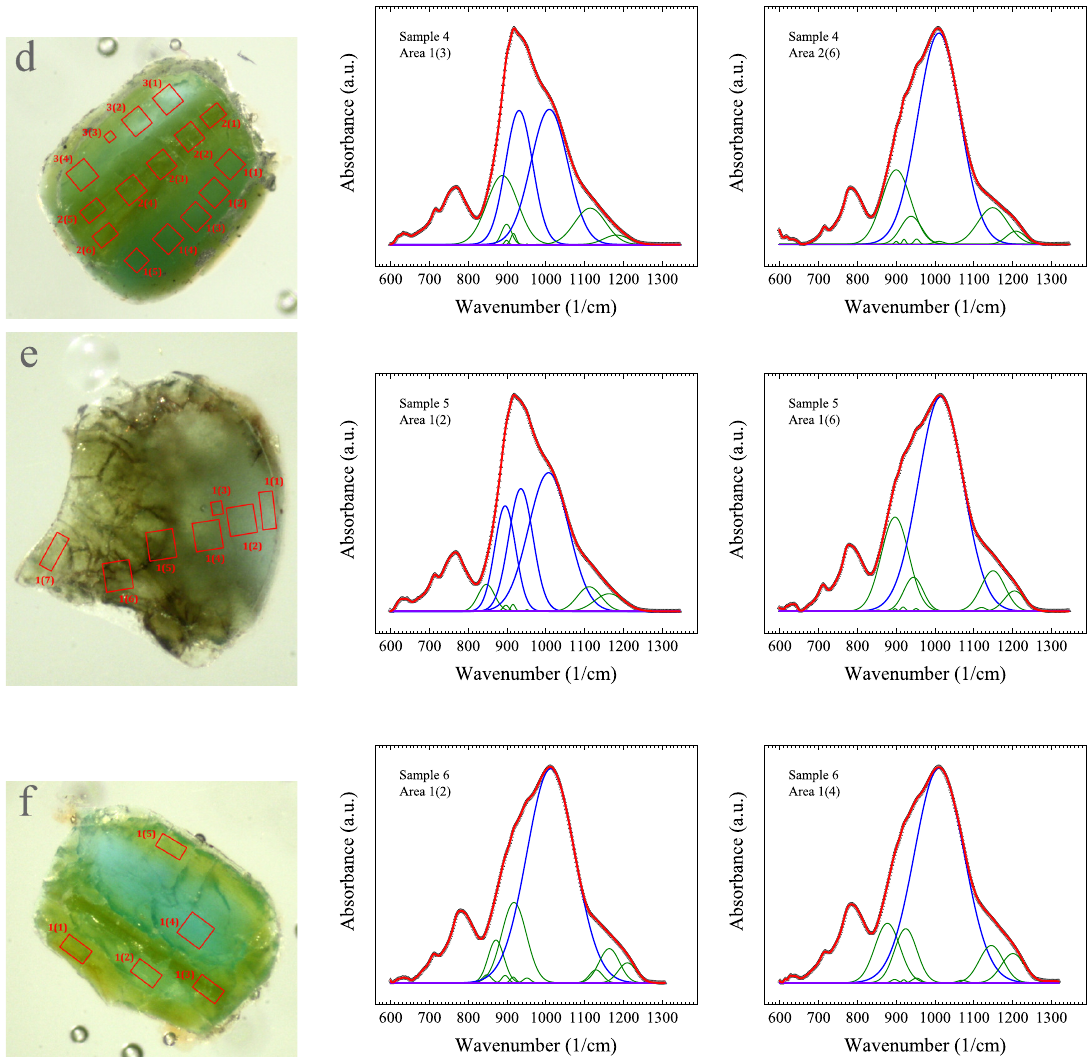}\\ 
		{\footnotesize  Figure~\ref{fig:Beads_FTIR} (continued)
		}
		\label{fig:Beads_FTIR1}	       
\end{minipage}
\end{figure}

\begin{figure}[th]
	\begin{minipage}[l]{1.2\textwidth}{\vspace{-3cm}}{\hspace{-1cm}}
				\includegraphics[scale=1]{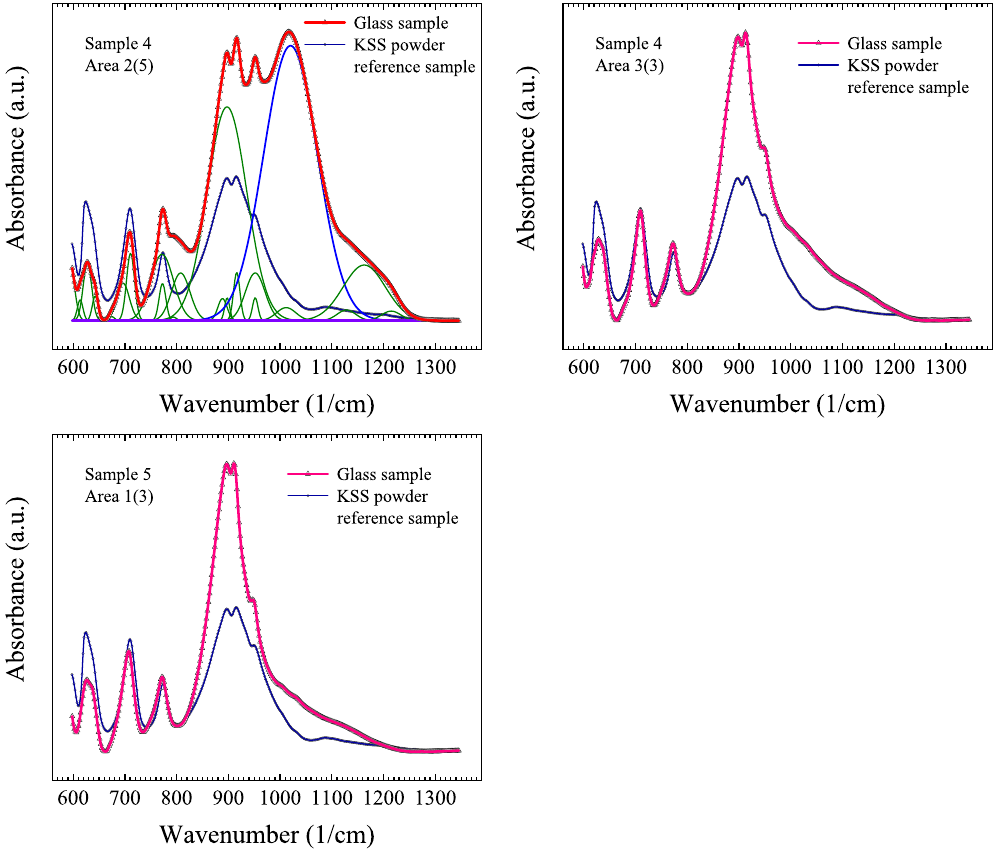}
		\caption{(Colour online)
			IR absorption spectra obtained at several areas of the semi-opaque turquoise glass bead samples containing a considerable amount of KSbOSiO$_4$:
			a larger area of glass encompassing numerous KSS inclusions (sample~4, area~2(5), Fig.~\ref{fig:Beads_FTIR}\,d)
			and
			two smaller areas comprising mainly colonies of KSS crystals (sample~4, area~3(3) and sample~5, area~1(3), Fig.~\ref{fig:Beads_FTIR}\,d,\,e);
			the spectra represent a superposition of glass and KSS absorption bands with the latter ones predominating in the spectra recorded at the smaller areas;
			the reference spectrum of KSS powder is given for the comparison in each panel by the royal blue line
			(for details on the KSS IR absorption spectrum, see 
			Online Resource~\ref{esm:fig_KPbSbO5});
			empty triangles and circles depict the experimental data,
			the pink lines show the spectra from areas with dominating absorption by KSS crystals,
			the olive and blue lines present peaks derived as a result of the spectrum deconvolution 
			(the blue line shows the strongest absorption peak),
			the red line is a cumulative fit curve derived from the peak analysis.
		} 
	\label{fig:Beads-KSS_FTIR}%
	\end{minipage}    
\end{figure}
\clearpage

\subsection{FTIR spectroscopy} \label{subsec:results_FTIR}

In Fig.~\ref{fig:Beads_FTIR}, examples of FTIR spectra obtained at undamaged and degraded regions of bead glass of the samples shown in Fig.~\ref{fig:Beads_samples} are presented.
(A complete set of FTIR spectra for all the marked areas is available in Online Resource~\ref{esm:fig_FTIR_All_Spectra}.)
The spectra are seen to exhibit the same peculiarities as the spectra of intact and strongly deteriorated beads presented in our previous work cited in Ref.~\cite{2020_Glas_depolimer}.
The strongest absorption bands are peaked around the wavenumbers of 910--920~cm$^{-1}$ in the spectra recorded at undamaged glass areas and around 990--1030~cm$^{-1}$ in the spectra obtained at degraded glass areas. 
Some spectra reach a maximum at about 950~cm$^{-1}$ (see Online Resource~\ref{esm:fig_FTIR_All_Spectra}), yet this should be assigned to the strong contribution of IR absorption by KSS crystals in the areas, at which these spectra have been obtained (for the KSS absorption lines, see Online Resource~\ref{esm:fig_KPbSbO5}).

The contribution of KSS to the main band is usually manifested as a sharp tip on its top or as oscillations on its left-hand slope.
The less intense band or a set of bands blue-shifted with respect to the main one (located at the wavenumbers below 840~cm$^{-1}$) should be assigned to a superposition of KSS and glass absorption lines \cite{RCEM-2018_vibrational_spectroscopy,Preservation_Cultural_Heritage-2018,GosNIIR_Conference,Beads_EAMC-2018}.%
\footnote{%
These long-wavelength vibrations will not be discussed in this article, but the corresponding KSS mid- and far-IR absorption bands are shown in Figs.~ESM~\ref{esm:fig_KPbSbO5}.3 to ESM~\ref{esm:fig_KPbSbO5}.5 and listed for reference in Tables~ESM~\ref{esm:fig_KPbSbO5}.1 and ESM~\ref{esm:fig_KPbSbO5}.2 (Online Resource~\ref{esm:fig_KPbSbO5}) nonetheless.
}

We have decomposed the spectra to Gaussian peaks to determine the strongest vibrations in accordance with peak intensities and integrals. 
The results of the deconvolution are presented in Fig.~\ref{fig:Beads_FTIR} and Online Resource~\ref{esm:fig_FTIR_All_Spectra}.%
\footnote{%
	All moderate and weak components derived from the spectra as a result of their deconvolution that are peaked at the wavenumbers below 840~cm$^{-1}$ have been excluded from the plots for all the spectra except for those exhibiting the dominating IR absorption by KSS crystals (Online Resource~\ref{esm:fig_KPbSbO5}, Figs.~ESM~\ref{esm:fig_KPbSbO5}.3 to ESM~\ref{esm:fig_KPbSbO5}.5, Tables~ESM~\ref{esm:fig_KPbSbO5}.1 and ESM~\ref{esm:fig_KPbSbO5}.2).
	The omitted peaks are assigned to vibrations both in KSS and in glass, yet their high number, rather low intensity and ambiguous data of peak analysis (often, several combinations with close $R^2$ values give the same cumulative fit curve) have made us refuse their presenting in the graphs to avoid confusion.
	Notice also, that they are not related to the issue discussed in the article.
}

It is easy to see that the spectra obtained at degraded glass areas are dominated by the only vibration band always peaked at the interval from approximately 1000 to nearly 1030~cm$^{-1}$, whereas there are no dominating peaks in the spectra obtained at intact glass areas but on the contrary several vibration peaks of close intensities and comparable integrals, which are distributed rather uniformly within the main band, are always present.

Fig.~\ref{fig:Beads-KSS_FTIR} demonstrates IR absorption spectra obtained at several glass areas that contain a considerable amount of KSS crystals clustered to large colonies.
The plot for sample 4, area 2(5) shows a superposition of bands assigned to vibrations in KSS and glass, with a peak at about 1020~cm$^{-1}$ predominating in the glass-related absorption, that is typical to degraded turquoise-bead glass.
The other two panels (sample 4, area 3(3) and sample 5, area 1(3)) represent the spectra that are completely dominated by the KSS absorption bands, with no glass-related IR absorption bands being recognized in them except for the far right-hand wing of the main band at the wavenumbers above 1000~cm$^{-1}$.    

We should emphasize in the conclusion of this section that there is a significant difference in the composition of IR absorption spectra of glass acquired at undamaged and degraded domains of beads, which has been verified at a rich data array.
The characteristic feature of the spectra from corroded glass domains is that all vibration bands in them are dominated by a single intense one peaked in the range from $\sim 1000$ to $\sim 1030$~cm$^{-1}$, whilst the spectra from intact glass domains always comprise a set of vibration bands with comparable peak areas that are distributed from $\sim 870$ to $\sim 1000$~cm$^{-1}$. 
In domains of moderately corroded glass, the IR spectra are often composed of a few bands of comparable area but the band peaked around 1000~cm$^{-1}$ always has the greatest area and intensity nevertheless (see Online Resource~\ref{esm:fig_FTIR_All_Spectra}). 

The above completely confirms the results previously reported by us in Ref.~\cite{2020_Glas_depolimer} for separate intact and corroded beads.

\subsection{EDX analysis}  \label{subsec:results_EDX}

\subsubsection{Elemental analysis}  \label{subsubsec:results_EDX_analysis}

EDX microanalysis was carried out at points of different preservation degree on all six samples shown in Fig.~\ref{fig:Beads_samples}.
The results are presented in Table~\ref{tab:EDS-1}.
The points of analysis were subdivided into three categories in accordance with their glass preservation state, namely, undamaged or slightly damaged blue glass site, moderately damaged or yellow glass one, and heavily damaged or dark glass one. 
One can see from Table~\ref{tab:EDS-1} that the content of alkali metals is considerably decreased in points located at moderately or heavily damaged sites compared to their content in points situated at undamaged or slightly damaged ones.
It is roughly 1.5 to 2 times lower for K and 3 to 4 times lower for Na at domains of corroded glass. 
The only exclusion is the point \#\,6~($h$), which reflects mainly the elemental composition of a large accumulation of KSS crystals seen as an extended white area among large cracks and marked with an arrow in Fig.~\ref{fig:Beads_samples}\,r.%
\footnote{%
	Remind that Sb is contained only in KSS (KSbSiO$_5$) inclusions in this kind of glass \cite{Yur_JAP,Beads_KSS_SPIE,KSS_Electron_microscopy}).
}
Below, we will discuss the cause of the escape of atoms of alkali metals from the degrading regions of glass.%
\footnote{%
We should note at this point that we failed to reveal this fact previously in Ref.~\cite{2020_Glas_depolimer} when the elemental analysis was made by means of X-ray fluorescence at individual beads or their separate fragments with a large spot of X-ray emission that covered a major part of a sample. 
Besides, the concentration of K atoms in glass of those samples likely significantly varied from sample to sample originally, at the moment of their production, that has veiled the loss of K owing to the glass corrosion because of the data scatter in the results of the elemental analysis of different samples.
}

Further, the relative content of Si atoms, on the contrary, is seen to be by approximately 20 to 30\,\% higher in moderately or heavily damaged domains than in undamaged or slightly damaged ones.

Finally, no correlation of the content of Ca, Cu, Sb, and Pb atoms with the glass degradation degree has been detected.

\subsubsection{Elemental mapping} \label{subsubsec:results_EDX_mapping}

\begin{figure*}[t]
\begin{minipage}[l]{1.2\textwidth}{\vspace{0cm}}
		\includegraphics[scale=1]{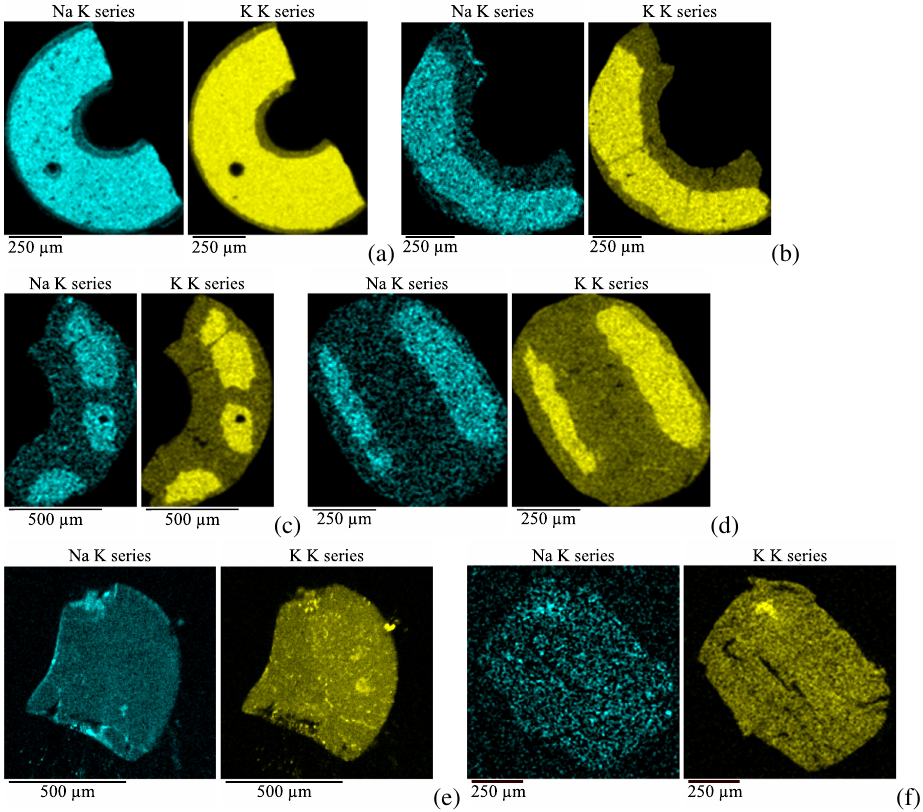}
	\caption{(Colour online) 
		EDX maps of Na and K distribution on the sections of semi-opaque turquoise glass beads:
		(a) sample~1,
		(b) sample~2,
		(c) sample~3,
		(d) sample~4,
		(e) sample~5,
		(f) sample~6. 
	}
	\label{fig:Beads_maps}	 
\end{minipage}      
\end{figure*}

\begin{figure*}[th]
		\begin{minipage}[l]{1.2\textwidth}{\vspace{0cm}}{\hspace{0cm}}
				\includegraphics[scale=1]{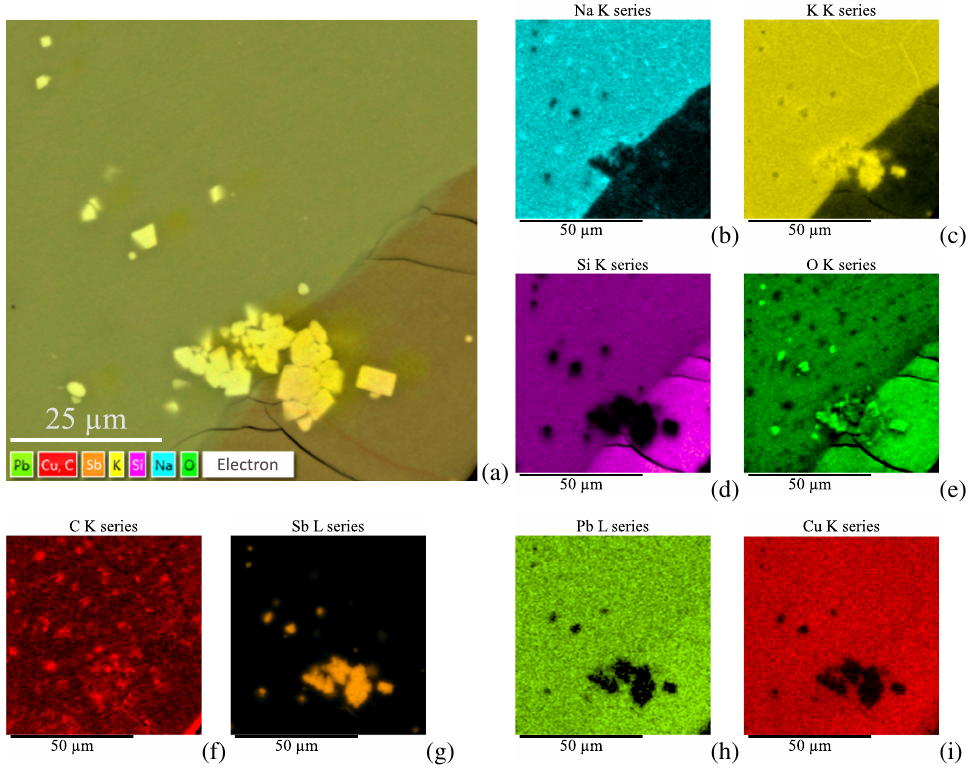}
		\end{minipage}
	\caption{(Colour online) 
		Chemical element distribution maps acquired in an area of the sample 1 around the core-shell interface
		near the bead hole:
		a layered map
		comprising X-ray maps superimposed on a 
		SEM BSE image~(a)
		and
		radiation intensity  
		maps for X-ray lines of
		Na~(b) and K~(c),
		Si~(d) and O~(e),
		C~(f) and Sb~(g),
		and
		Pb~(h) and Cu~(i);
		bright inclusions seen in the panel (a) are KSS crystallites.
	}
	\label{fig:Beads_maps_interface}	 
\end{figure*}

Elemental mapping of the samples has shown the same peculiarities in the distribution of K and Na (Fig.~\ref{fig:Beads_maps}) \cite{RCEM-2018_vibrational_spectroscopy,Preservation_Cultural_Heritage-2018,Grabar-2018}.
The number density of alkali metal atoms is seen to be significantly higher in undamaged glass domains than in damaged ones.
In addition, the distributions of K and Na are seen to exactly coincide in all the studied samples.
Nearly intact sample~1 (Fig.~\ref{fig:Beads_maps}\,a) demonstrates practically uniform distribution of alkali metals throughout its volume except for the region that is immediately adjacent to its hole, whilst the most corroded samples~5 and 6 (Fig.~\ref{fig:Beads_maps}\,e,\,f) have mostly lost K and Na: K is mainly seen in KSS crystals, whereas Na is observed as small accumulations, which, as follows from Fig.~\ref{fig:Beads_maps_interface}, are formed by its carbonates.
We would like to attract the readers' attention to the fact that domains depleted with K and Na are present not only at sample surfaces (or in their shells) but also in the internal regions (in the cores) that is clearly seen in Fig.~\ref{fig:Beads_maps}\,c,\,d, while alkali metal-rich domains are often seen close to the surface.

Let us proceed to the maps of other elements.
We present as an example a detailed analysis of the sample~1 in Fig.~\ref{fig:Beads_maps_interface}.
Its corroded near-hole domain and intact bulk region enable elemental profiling aimed to determining fine features of chemical elements distribution at the core-shell interface (i.e. at the interface between damaged and intact glass).

In Fig.~\ref{fig:Beads_maps_interface}, light spots on the K, O and Sb distribution maps (the panels~c,~e and~g) and dark spots on the maps of the Na, Si, Pb and Cu distribution (the panels~b, d, h and i) corresponding to them are due to the KSS crystals that are seen in the layered map.
Na and C maps are highly correlated indicating the presence of inclusions of sodium carbonates in glass.
Locations of bright spots on the maps of the Na and C distribution also coincide with dark spots on the map of the O distribution (Fig.~\ref{fig:Beads_maps_interface}\,b,\,e,\,f).
Clusters of soda ash (or natron) have likely formed owing to a raw material non-uniformly distributed during the glass making, which then has remained mainly accumulated in the inclusions.

The electron (BSE) image given as a background layer on the layered map presented in Fig.~\ref{fig:Beads_maps_interface}\,a demonstrates that the intact glass of the core is heavier than the degraded one of the near-hole shell layer (a broad darker stripe in the lower right corner of the image).
K and Na distribution maps (Fig.~\ref{fig:Beads_maps_interface}\,b,\,c) show a significantly decreased content of these elements in the near-hole shell (corroded) layer (the latter sector is seen nearly black on the maps), whereas Si and O distribution maps (Fig.~\ref{fig:Beads_maps_interface}\,d,\,e), on the contrary, demonstrate a significantly increased number density of these elements in this site. 

According to the maps presented in Fig.~\ref{fig:Beads_maps_interface}\,h,\,i, Pb and Cu are distributed quite uniformly; minor alterations of the X-ray emission intensity at Pb\,L$_{\alpha}$ and Cu\,K$_{\alpha}$ bands in the near-hole layer may be caused by some other reasons, e.g., by a denser silicate matrix, rather than an increased content of Pb and Cu.
Thus, the visible minor growth of their content in this site is not entirely reliable and apparently should be disregarded. 

It is important to note also that mapping of C content has revealed a thin near-surface layer of glass in the bead hole that is enriched with carbon (a bright read stripe in the right lower corner of Fig.~\ref{fig:Beads_maps_interface}\,f) that likely migrated there from the polishing powder, which filled bead holes during tumbling at elevated temperature and consisted of clay, chalk and charcoal \cite{KSS_Electron_microscopy,Yurova_large}. 
Remark also that Ca is distributed inhomogeneously in the glass bulk that is reflected in the data given in Table~\ref{tab:EDS-1} (X-ray maps of Ca distribution are not presented in this article).
We failed to observe any difference in the distribution of Ca in degraded and undamaged domains. 

\clearpage

\subsubsection{Elemental profiles}   \label{subsubsec:results_EDX_profiles}

\begin{figure*}[t]
\begin{minipage}[l]{1.2\textwidth}{\vspace{0cm}}
		\includegraphics[scale=1]{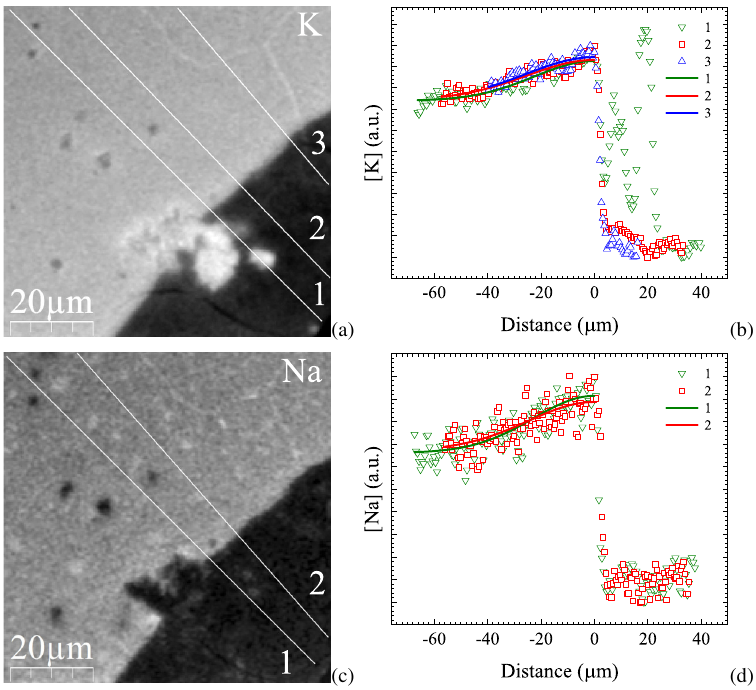}
\end{minipage}
	\caption{(Colour online) 
		K (a, b) and Na (c, d) concentration profiles (b, d) along lines crossing the core-shell interface near the bead hole (a, c) in the sample~1; 0 of abscissa axis is set at the interface, the axis is directed to the orifice; open symbols present experimental data, solid lines are diffusion profiles obtained by data fitting; the numbers in the graphs (b, d) correspond to the numbers of the lines on the respective X-ray maps (a, c). 
	}
	\label{fig:Beads_K-Na_profiles}	       
\end{figure*}

\begin{figure*}[th]
\begin{minipage}[l]{1.2\textwidth}{\vspace{0cm}}
		\includegraphics[scale=1]{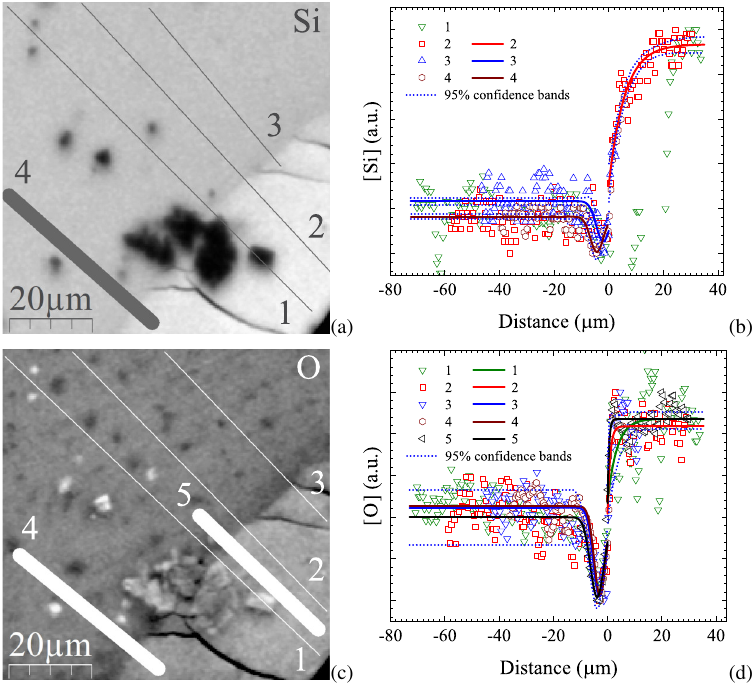}
\end{minipage}
	\caption{(Colour online) 
		Si (a, b) and O (c, d) concentration profiles along lines crossing the core-shell interface near the bead hole (a, c) in the sample~1; 0 of abscissa axis corresponds to the interface, the axis is directed to the orifice; open symbols present experimental data, solid lines show concentration profiles obtained by data fitting, and dotted lines are 95\% confidence bands; the numbers in the graphs (b, d) correspond to the numbers of the lines on the respective X-ray maps~(a, c); stripes 4 and 5 display the width, within which the profile data are averaged. 
	}
	\label{fig:Beads_Si-O_profiles}	       
\end{figure*}

In order to deeper investigate the distribution features of alkali metals (K and Na), and the silicate-matrix forming elements (Si and O) in the vicinity of the core-shell interface, we have plotted their distribution line profiles across the interface (Figs.~\ref{fig:Beads_K-Na_profiles} and \ref{fig:Beads_Si-O_profiles}); the profiles have been derived from the corresponding maps presented in Fig.~\ref{fig:Beads_maps_interface} (the panels~b to e).

Both K and Na are seen to exhibit similar peculiarities of their distribution (Fig.~\ref{fig:Beads_K-Na_profiles}).
Being nearly constant in the intact core, their concentrations start growing as Gaussian functions at the distance of about 55 to 60~{\textmu}m from the interface reaching maximum directly at the interface.
Then, in the shell layer of degraded glass, they steeply fall down within the length of about 5~{\textmu}m from the interface and become constant again, but at a lower level, in that layer. 
 
The profiles of Si and O are also similar but strongly differ from those of K and Na (Fig.~\ref{fig:Beads_Si-O_profiles}). 
They are also constant in the core, yet begin falling down at the distance of about 10~{\textmu}m from the core-shell interface reaching minimum at the distance of 2 to 5~{\textmu}m from it. 
Then, at the proximity of the interface, the concentrations start to rapidly increase until saturate at a higher level in the shell layer at the distance of about 5 to 10~{\textmu}m from the interface.

Note that it is hard to estimate the above lengths more accurately due to the spatial noise caused by the inhomogeneity of the elemental composition that introduces a considerable scatter in the data to be fitted.
Nevertheless, the general features can be derived from the presented graphs.
Firstly, alkali metals leave the corroded shell layer to accumulate at the interface in the intact core.
Secondly, the number density of Si and O in the shell is greater than in the core, i.e. the glass silicate matrix is denser in the shell.
Finally, there are two sub-interfacial layer in the intact core region exhibiting special properties: a 10-{\textmu}m thick layer, in which glass is less dense than in the surrounding bulk and especially in the shell, and $\sim50$-{\textmu}m thick one, in which an increased concentration of alkali metal atoms is observed;
the latter demonstrates Gaussian impurity concentration profiles.

\clearpage

\clearpage

\subsection{Glass degradation during sample storage} \label{subsec:sample_storage} 

\subsubsection{Glass cracking and redistribution of chemical elements} \label{subsubsec:sample_storage_cracks}

In the lab, bead samples are stored in conditions close to those of keeping in museums.
However, gradual glass degradation is often observed in samples of unstable beads.
This is especially clearly seen in samples of heavily degraded turquoise beads.
Sometimes, turquoise beads at the final stages of corrosion crumble to particles for the time of several years.
Usually we see the consequences of this process when visually check up the conservation state of the previously examined samples.
Now we present the results of the systematic study of this phenomenon using the sample~4 as an example.

\begin{figure*}[th]
	\begin{minipage}[l]{1.2\textwidth}{\vspace{0cm}}{\hspace{-1cm}}
			\includegraphics[scale=.9]{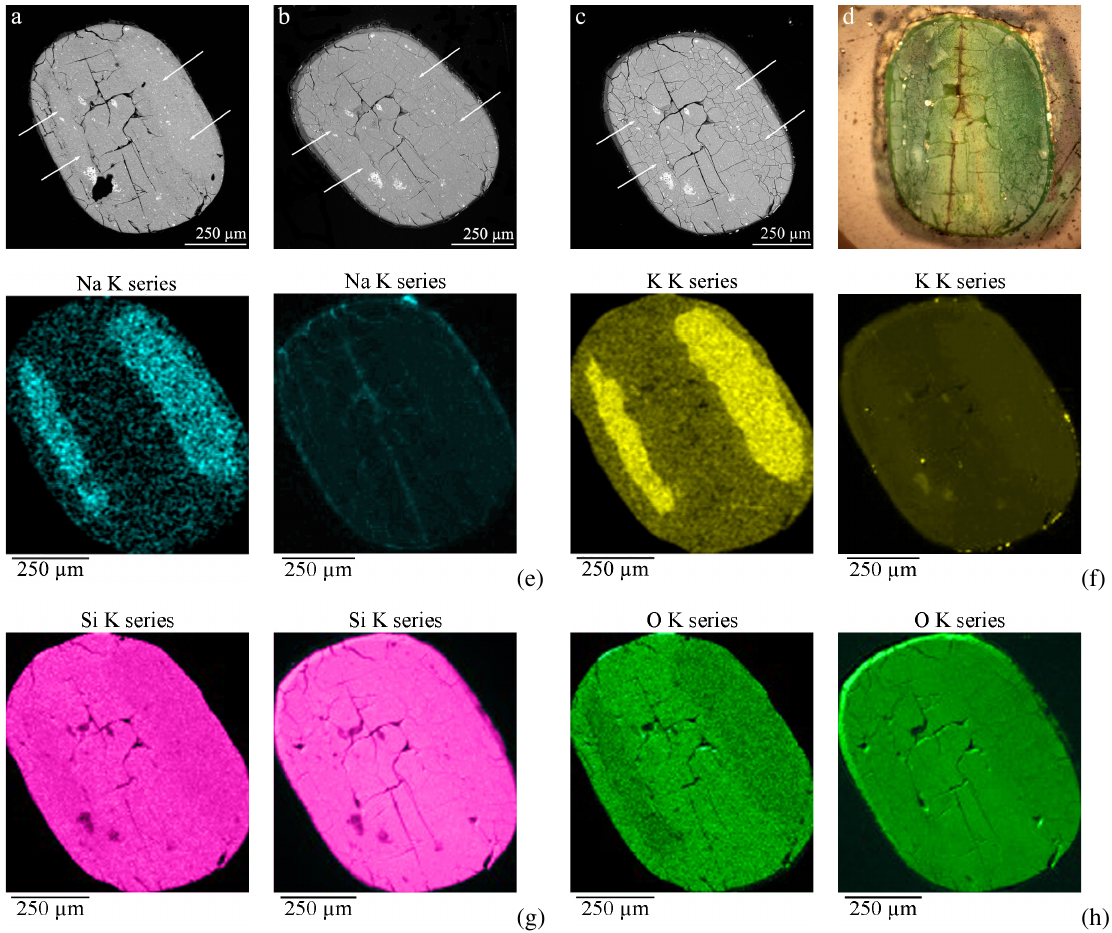}
	\caption{(Colour online)
	Glass degradation during sample storage in the lab: SEM BSE images of the sample~3 obtained on 2017.11.21~(a), 2020.08.10~(b) and 2021.07.08~(c); numerous newly arisen cracks are seen in the images (b) and (c) [compare the the domains shown with arrows in the panels (a), (b) and (c)]; heavier-glass domains (lighter grey areas) are seen both in the right-hand and left-hand areas of the bead image (a) [shown with arrows], which coincide with the regions of an increased content of Na and K in Fig.\,\ref{fig:Beads_maps}\,d and the panels (e) and (f), that disappear in the images (b) and (c); a micro photograph of the sample~3 obtained on 2021.07.06~(d) that also demonstrates a network of new cracks (compare with the photograph of 2018.07.23 shown in Fig.~\ref{fig:Beads_samples}\,j); pairs of EDX maps for Na (e), K (f), Si (g) and O (h) K$_{\alpha}$ bands obtained on 2017.11.21~(left) and 2021.07.08 (right).
	}
	\label{fig:Sample_degradation}
\end{minipage}		       
\end{figure*}

Fig.~\ref{fig:Sample_degradation} demonstrates the process of such degradation observed in the sample~4 using SEM and LM.
For the period from 2017.11.21 (the panel~a) to 2021.07.08 (the panels~c,\,d), numerous new cracks have appeared in the bead section.
They are seen to arise in the glass domains that were nearly uncracked previously.
Additionally, e.g., the right-hand (previously uncracked) part of the sample is seen to be paler in the earlier image than the same region in the later ones (Fig.~\ref{fig:Sample_degradation}\,a--c), which means that glass was originally heavier in this domain and then has become lighter owing to the outmigration of potassium from this region (see also Fig.\,\ref{fig:Beads_maps}\,d).
The similar process has happened in the area to the left of the heavily cracked central area.
(Both these areas are shown with white arrows in Fig.~\ref{fig:Sample_degradation}.)

The concentrations of Na and K, as expected, are seen to have drastically fallen in these areas (Fig.~\ref{fig:Sample_degradation}\,e,\,f) and practically equalized throughout the analysed layer.
The number densities of Si and O atoms, in turn, have grown in these regions (Fig.~\ref{fig:Sample_degradation}\,g,\,h) and virtually reached those in the heavily cracked central region.

We associate these closely connected phenomena with the internal stress, which is present in glass even after the section preparation, causing the out-diffusion of alkali metals from the near-surface layer accompanied with slow fracturing of glass. 
This process eventually results in the depletion of the near-surface areas of the alkali metal-rich domains (Fig.\,\ref{fig:Beads_maps}) and the equalisation of the concentrations of both K and Na throughout a sample near-surface layer.%
\footnote{%
X-ray fluorescence show that the excess concentration of Na and K had remains deeper in glass, however. 	
} 
The increase in the density of the silicate matrix of glass evidences in favour of alterations in its structure.

\subsubsection{FTIR analysis} \label{subsubsec:sample_storage_FTIR} 

\begin{figure*}[th]
	\begin{minipage}[l]{1.2\textwidth}{\vspace{0cm}}{\hspace{-2cm}}
		\includegraphics[scale=1]{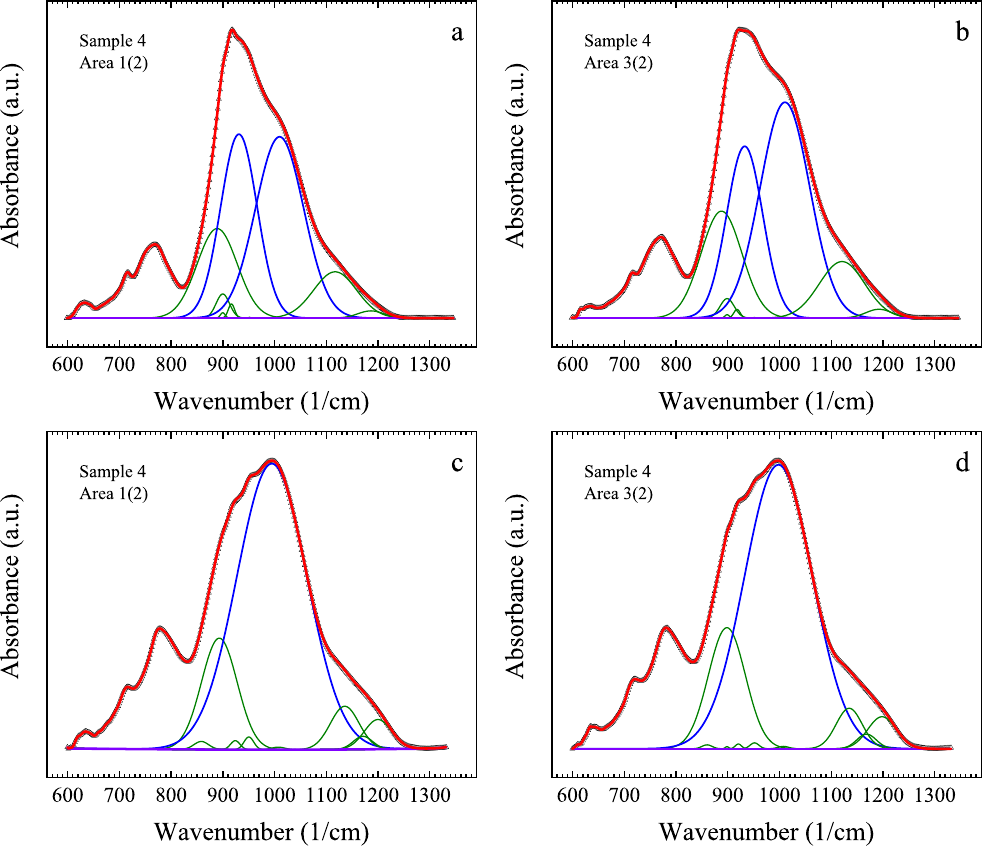}
	\end{minipage}
	\caption{(Colour online) 
		IR absorption spectra recorded at the same areas of the sample~4 at the domains of preserved glass on 2017.12.07 (a, b) soon after the sample preparation and on 2021.07.22 (c, d) after the glass degradation (Fig.~\ref{fig:Sample_degradation}):
		(a, c) area 1(2)
		and
		(b, d) area 3(2);
		the designations are the same as in Fig.~\ref{fig:Beads_FTIR};
		to compare more FTIR spectra of this sample before and after the degradation, see Figs.~ESM~\ref{esm:fig_FTIR_All_Spectra}.4 and~ESM~\ref{esm:fig_FTIR_All_Spectra}.7 in Online Resource~\ref{esm:fig_FTIR_All_Spectra}.
	}
	\label{fig:Sample_degradation_FTIR}	       
\end{figure*}

We have re-examined the IR absorption in the sample~4 having recorded the spectra after its degradation described in the previous Section at the same areas, at which it was done shortly after section polishing.
Fig.~\ref{fig:Sample_degradation_FTIR} demonstrates the IR absorption spectra obtained at the domains of preserved glass on 2017.12.07 (the panels~a, b) soon after the sample preparation and on 2021.07.22 (the panels~c, d) after the glass degradation (see also Figs.~ESM~\ref{esm:fig_FTIR_All_Spectra}.4 and~ESM~\ref{esm:fig_FTIR_All_Spectra}.7 in Online Resource~\ref{esm:fig_FTIR_All_Spectra}).
Right after the sample preparation, the spectra main bands, as shown in Section~\ref{subsec:results_FTIR}, reached their maxima at the wavenumbers of $\sim$\,910--920~cm$^{-1}$, yet their maxima are seen to have shifted to $\sim$\,1000~cm$^{-1}$ because of the glass degradation, and the spectra have become very alike to those previously obtained at the degraded areas (Fig.~\ref{fig:Beads_FTIR}).

\clearpage
\clearpage

\clearpage
\clearpage

The peak analysis, as expected, has yielded the same results as in Section~\ref{subsec:results_FTIR}. 
We have revealed the only dominating peak at approximately 1000~cm$^{-1}$ in the spectra obtained at the domains of newly degraded glass (Fig.~\ref{fig:Sample_degradation_FTIR}\,c,\,d).
This evidences that both the process of the long-term glass corrosion and that of the gradual glass degradation after polishing comprise glass depolymerization.

\subsection{Effect of annealing: artificial ageing}  \label{subsec:results_aging}

\subsubsection{Shell formation and fracture}  \label{subsubsec:results_aging_structure}

\begin{figure*}[th]
	\begin{minipage}[l]{1.2\textwidth}{\vspace{-2cm}}
		\includegraphics[scale=.9]{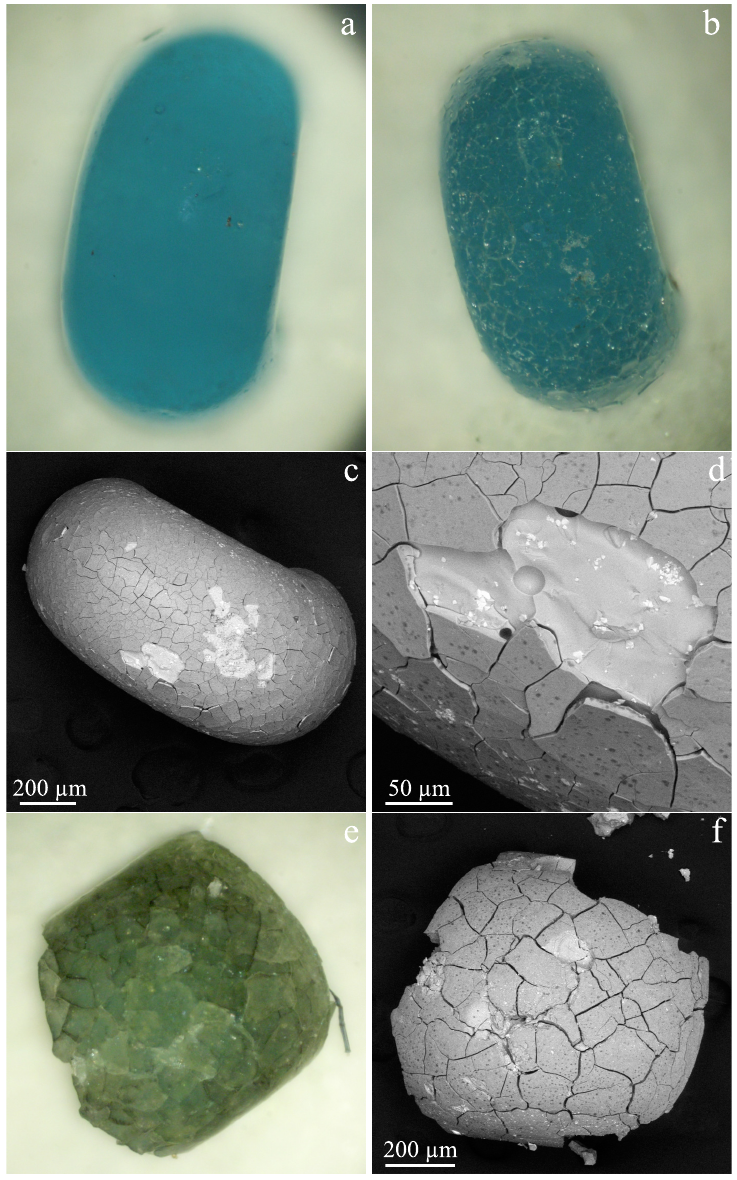}
	\caption{(Colour online) 
		Effect of annealing at {300\textcelsius} for 15~minutes in the atmospheric air on turquoise glass beads: 
		light micro images of an intact bead before (a) and after (b) the heat treatment;
		SEM BSE images of the same bead after the annealing (c, d): 
		a crust consisting of glass flakes separated by a network of cracks is seen as well as `windows' to the bead core opened due to flake shedding;
		light (e) and SEM BSE (f) images of a bead at the late stage of corrosion \cite{Yuryev_JOPT,Yur_JAP}
		before (e) and after (f) the annealing: the glass shell structure is similar to that seen in the panels (c) and (d).
		 	}
	\label{fig:Artificial_aging}
\end{minipage}	       
\end{figure*}

Thermal treatment of intact beads at the temperature of {300\textcelsius} for 15~minutes in the atmospheric air resulted in the formation of a fractured crust on the sample surface (Fig.~\ref{fig:Artificial_aging}\,a--d).
The crust resembles the shell, which is often observed in heavily degraded beads, and varies in the thickness from nearly 10 to about 20~{\textmu}m.

Heavily degraded beads at the late phase of corrosion (Fig.~\ref{fig:Artificial_aging}\,e,\,f) visually practically did not change as a result of the annealing, yet they became more fragile and started rapid crumbling (Fig.~\ref{fig:Artificial_aging}\,f).
However, they demonstrate the same crust structure as intact ones after the heat treatment:
the shell is fractured and consists of glass flakes separated by a network of deep cracks, which reach the bead core.

The glass shell is seen to shed flakes that opens the access to glass of the core domain for direct analysis.

This artificial ageing process simulates the long-term corrosion demonstrating how the shell might be formed on turquoise beads.

\clearpage
\clearpage

\begin{figure*}[th]
	\begin{minipage}[l]{1.2\textwidth}{\vspace{0cm}}{\hspace{0cm}}
		\includegraphics[scale=.9]{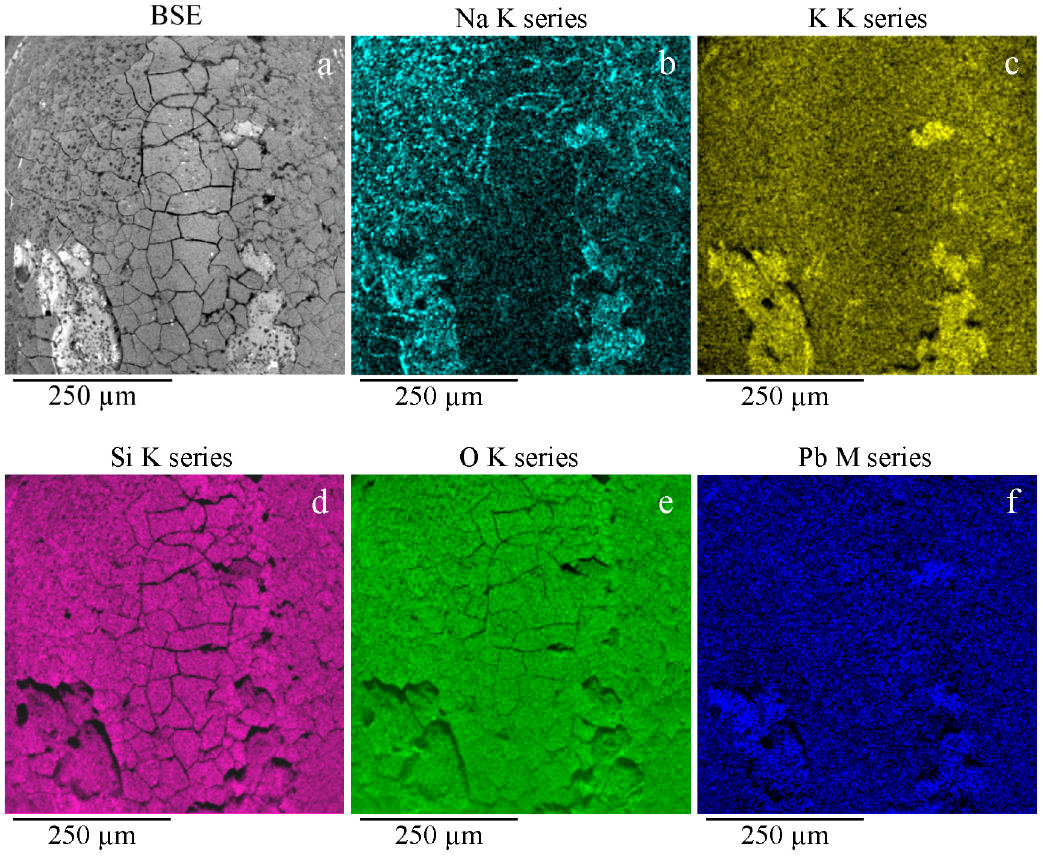}
	\end{minipage}
	\caption{(Colour online) 
		Images of a glass shell area on the heat treated initially intact bead (Fig.~\ref{fig:Artificial_aging}\,b--d):
		a SEM BSE image~(a) showing cracks, scales and empty places appeared because of scale exfoliating (`windows'),
		and
		maps of sodium, potassium, silicon, oxygen and lead distribution (b\,--\,f).
	}
	\label{fig:Artificial_aging_EDX}	       
\end{figure*}
\clearpage

\subsubsection{Redistribution of chemical elements} \label{subsubsec:results_aging_maps}

\begin{figure*}[th]
	\begin{minipage}[l]{1.2\textwidth}{\vspace{0cm}}{\hspace{-0cm}}
		\includegraphics[scale=1]{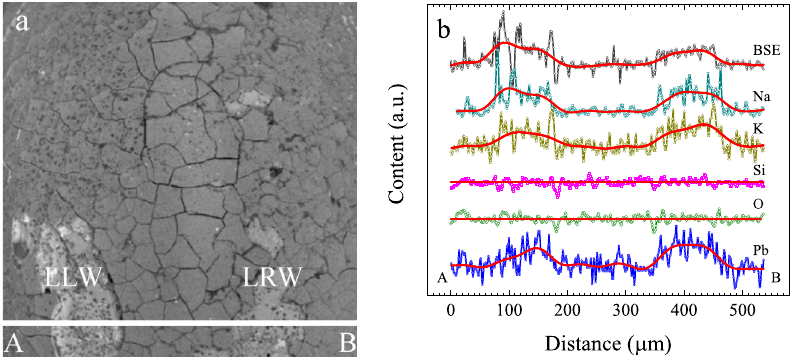}
	\end{minipage}
	\caption{(Colour online)
		Line scan profiles recorded from an initially intact turquoise bead sample after the annealing (Fig.~\ref{fig:Artificial_aging}\,b--d) 
		along the line segment [AB] shown in the SEM BSE image (a):
		the line crosses the lower left and lower right windows (LLW and LRW) to the core domain;
		the data for each profile have been averaged within the [AB] line width;
		the profiles (b) are presented for the SEM BSE image brightness and for the intensities of X-ray bands of Na, K, Si, O and Pb mapped in Fig.~\ref{fig:Artificial_aging_EDX}; 
		a type of the scanned signal is indicated in the plot (b) next to the corresponding curve;
		red lines obtained by fast Fourier transform filtering are given as a guide to the eye.
	}
	\label{fig:Artificial_aging_profiles}	       
\end{figure*}

Chemical elements are seen to redistribute in glass during the annealing (Figs. \ref{fig:Artificial_aging}\,c,\,d,\,f, \ref{fig:Artificial_aging_EDX} and~\ref{fig:Artificial_aging_profiles}).
In Figs.~\ref{fig:Artificial_aging}\,c,\,d,\,f and~\ref{fig:Artificial_aging_EDX}\,a, one can see that glass of the shell has become lighter than that of the core due to the heat treatment.
The brightness profile of the BSE image (Fig.~\ref{fig:Artificial_aging_profiles}\,b) also shows that the core glass is somewhat heavier than the shell one.

The maps of Na and K distribution and the line scan profiles (Figs.~\ref{fig:Artificial_aging_EDX}\,b,\,c and~\ref{fig:Artificial_aging_profiles}\,b) demonstrate a less content of these elements in the shell than in the core that corresponds with Fig.~\ref{fig:Beads_Si-O_profiles};
Pb concentration is also higher in the core than in the shell (Figs.~\ref{fig:Artificial_aging_EDX}\,f and~\ref{fig:Artificial_aging_profiles}\,b).
As distinct from Fig.~\ref{fig:Beads_Si-O_profiles}, Si and O content visually does not differ in the core and the shell (Figs.~\ref{fig:Artificial_aging_EDX}\,d,\,e and~\ref{fig:Artificial_aging_profiles}\,b).

Semi-quantitative elemental analysis data gathered in Table~\ref{tab:EDS-2} show that Na content is about 3.4--3.7 times lower in the shell than in the core whereas the content of K is only about 2 times higher in the core than in the shell. 
Si content is somewhat (by $\sim 20$\,\%) higher in the shell than in the core that is not detected in the X-ray map, however, as well as in the line scan profile since the spatial noise is too high in it (Figs.~\ref{fig:Artificial_aging_EDX}\,d and~\ref{fig:Artificial_aging_profiles}\,b).
The difference in Si content values was obtained perhaps because the shell was analysed by integrating the data over the area around the centre of the image, whereas the data of the core analysis were integrated over the windows; glass might originally somewhat differ in these areas.
(Nevertheless, this value agrees with the data of elemental analysis presented in Section~\ref{subsubsec:results_EDX_analysis}.)
Cu content approximately coincides in the core and the shell.
Sb, as mentioned above, is present only in KSS crystals, which are fused into glass, and its content in Table~\ref{tab:EDS-2} reflects only a number of crystals that got into the area of analysis.
Pb content is approximately 1.3 times higher in the core than in the shell that corresponds with the map in Figs.~\ref{fig:Artificial_aging_EDX}\,f and the profile in Fig.~\ref{fig:Artificial_aging_profiles}\,b.
Finally, as mentioned above (Section~\ref{subsubsec:results_EDX_mapping}), Ca atoms are distributed non-uniformly in this kind of glass, thus, the tabulated data on the Ca content mainly reflect its local fluctuations.

\clearpage

\subsubsection{Glass depolymerization}  \label{subsubsec:results_aging_depolymerization}

\begin{figure*}[th]
	\begin{minipage}[l]{1.2\textwidth}{\vspace{0cm}}{\hspace{-2cm}}
		\includegraphics[scale=1]{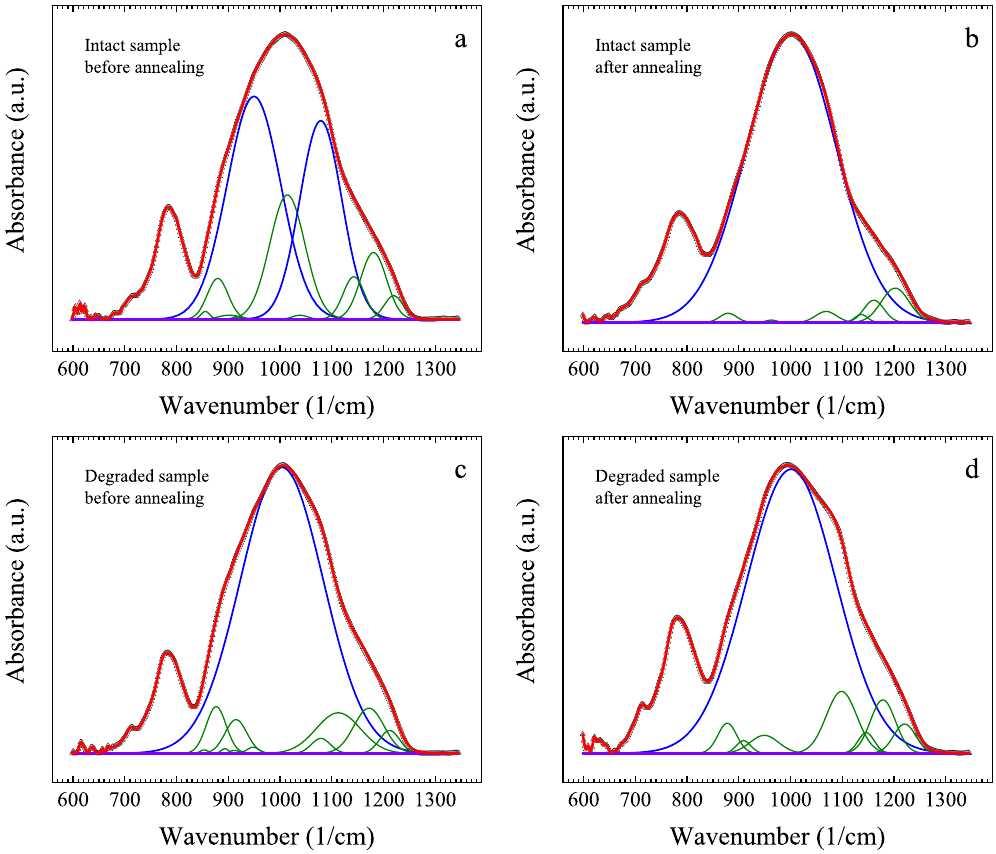}
	\end{minipage}
	\caption{(Colour online)
		Typical FTIR spectra recorded at intact (a, b) and degraded (c, d) turquoise bead samples before (a, c) and after (b, d) the annealing (Fig.~\ref{fig:Artificial_aging}); 
		the designations are the same as in Fig.~\ref{fig:Beads_FTIR}.
	}
	\label{fig:Artificial_aging_FTIR}	       
\end{figure*}

We have recorded FTIR spectra of the annealed beads and found that glass of the originally intact beads had depolymerized because of the heat treatment, whilst glass of the strongly degraded ones has not demonstrated any changes (Fig.~\ref{fig:Artificial_aging_FTIR}).
Remind that the studied samples, as distinct from those investigated in the above Sections, were not thin sections but untreated bead fragments, and all measurements were made at points on the bead shell in this Section. 

As it is seen in the spectrum obtained at the intact bead (Fig.~\ref{fig:Artificial_aging_FTIR}\,a), despite that the main absorption band peaks at about 1010~cm$^{-1}$, its analysis yields two intense components and several less intense ones that corresponds well with the above data for undamaged glass domains presented in Section~\ref{subsec:results_FTIR}. 
The intense bands peak at about 950 and 1080~cm$^{-1}$; 
the somewhat less intense one peaks at about 1010--1015~cm$^{-1}$.
After the heat treatment, the glass structure has changed (Fig.~\ref{fig:Artificial_aging_FTIR}\,b). 
Now, the main IR absorption band is seen to be composed of the only broad and very intense vibrational line, which peaks at about 1010~cm$^{-1}$, and several negligible ones in part assigned to KSS that completely agrees with the data for damaged glass domains presented above (Fig.~\ref{fig:Beads_FTIR}).

As expected based on the above results (Fig.~\ref{fig:Beads_FTIR}), the strongly degraded bead has demonstrated the main absorption band composed of the predominant vibrational line that has peaked at about 1000~cm$^{-1}$ and several faint ones both originally and after the thermal processing (Fig.~\ref{fig:Artificial_aging_FTIR}\,c,\,d). 
Glass of this sample obviously was depolymerized before the processing and has remained equally depolymerized after the treatment.

\clearpage
\clearpage

\section{Discussion}
\label{sec:discuss}

\subsection{Glass alteration} 
\label{subsec:discuss_alter}

\subsubsection{Glass depolymerization under stress} 
\label{subsubsec:discuss_depolymer}

As shown in Section~\ref{subsec:results_FTIR}, glass has decreased its polymerization degree in degraded domains and remained polymerized to much higher extent keeping its silicate unit composition comprising a set of radical species. 
We observe the same phenomenon as that we previously reported in Ref.~\cite{2020_Glas_depolimer} when studied separate corroded and intact glass beads. 
Therefore, the observed difference in the polymerization degree of glass degraded to different extent cannot be explained by the difference of silicate unit composition of the initial glass but only by the long-term process of its gradual destruction. 

Let us consider the glass structure evolution in terms of changes in a set of units composing it.

Since the number of vibrations detected in the areas of undamaged glass is always high and at least two (or sometimes three) of them demonstrate high intensities and close peak areas, we assume that several silicate units noticeably contribute to the glass structure in this case.
Based on Refs.~\cite{glass_units} and \cite{MOCIOIU-2013}, we tentatively assign the observed vibrations as follows: 
we assign the vibration bands peaked at $\sim$850, $\sim$870 to $\sim$880 to isolated tetrahedra (Q$^0$, SiO$_4^{4-}$), and  
those at $\sim$900$~cm^{-1}$ to paired ones (Q$^1$, Si$_2$O$_7^{6-}$); 
peaks at $\sim 950$~cm$^{-1}$ (and probably peaks at $\sim 930$~cm$^{-1}$) can be assigned to rings (Q$^2$, Si$_6$O$_{18}^{12-}$) \cite{glass_units}, 
yet they may also be related to chains (Q$^2$, Si$_2$O$_7^{6-}$) \cite{MOCIOIU-2013}; 
vibration bands peaked at $\sim$1000~cm$^{-1}$ to $\sim 1030$~cm$^{-1}$ may be assigned to chains (Q$^2$, Si$_2$O$_7^{6-}$) and probably double chains (Q$^2$\,+\,Q$^3$, Si$_4$O$_{11}^{6-}$) \cite{glass_units}, 
although the former attribution seems to be preferable since, according to Ref.~\cite{MOCIOIU-2013}, absorption band at 1006 to 1018~cm$^{-1}$ is characteristic to chains (Q$^2$) when the non-bridging oxygen atoms are bonded to both alkali metal cations and Pb$^{2+}$ ions that highly likely takes place in the studied glass; 
vibration bands peaked at $\sim$1110 to $\sim$1140~cm$^{-1}$ may be assigned to sheets (Q$^3$, Si$_2$O$_5^{2-}$) \cite{MOCIOIU-2013}; 
framework (Q$^4$, SiO$_2$) is perhaps responsible for the bands that peak at $\sim$1180 to nearly $1200$~cm$^{-1}$. 
We realize that this attribution is very approximate and somewhat speculative.
The provisional estimates of band maxima given in Refs.~\cite{glass_units,MOCIOIU-2013} are quite close to but somewhat differ from the peak position values derived by us and we endeavour to establish some correspondence with the former ones. 

Additionally, some peaks in the glass IR spectra are superimposed on those of potassium-antimony silicate crystals (see Online Resource~\ref{esm:fig_KPbSbO5}, Tables~ESM~\ref{esm:fig_KPbSbO5}.1 and ESM~\ref{esm:fig_KPbSbO5}.2) that also somewhat blurs the overall picture.
Besides, as follows from Section~\ref{subsec:stress}, the analysed glass is under the internal stress, which also should have a strong effect on the vibration-peak positions.
Remark, that the stress is distributed non-uniformly throughout a bead volume that introduces an additional uncertainty in the assignment of IR absorption peaks. 

Thus, all those factors considerably affect the data interpretation and interfere with the reliable peak assignment. 
However, we can certainly conclude that glass is composed of a broad spectrum of silicate species including Q$^3$ and Q$^4$ ones in intact domains, whereas in degraded domains, it mainly consists of the only type of silicate units, namely, of chains (Q$^2$), with a minor addition of other species.
In domains of moderately corroded glass, it is often composed of a narrower set of silicate species than it comprises in undamaged ones, with Q$^2$ chains being the main units.
Thus, we can state that the more destructed bead glass is, the more depolymerized it is.

We should remark at this point that the growth of Q$^2$ silicate unit content in expense of Q$^0$ and Q$^1$ ones is doubtful even under the effect of stress. The decay of highly arranged structures such as Q$^3$ and Q$^4$ appears to be a much more realistic scenario.
A physical mechanism of the glass depolymerization under the effect of internal stress was proposed in Refs.~\cite{RSF_16-18-10366/2018} and~\cite{2020_Glas_depolimer}.
We suggest that elementary acts of the glass depolymerization process are breaking of loaded interatomic bonds in silicate units by local fluctuations of the vibrational energy (thermal fluctuations \cite{Atomic-Level_Fracture_Solids}).
If an interatomic bond is under load, the energy barrier to its breaking decreases; 
as a result, the probability that a local fluctuation of the vibrational energy of this bond emerging at some moment will overcome that barrier increases; 
therefore, the likelihood of the bond splitting also increases. 
With the growth of a bond load, the barrier for its breakage lowers; 
therefore, with growing stress, the frequency of appearance as well as the mean density of bond vibrational energy local fluctuations, which could overcome the breaking barrier, increase in the bulk of glass.
Thus, the higher the internal stress, the more frequently the bonds split and the higher the glass depolymerization rate.  
In such a way, the depolymerization facilitates the stress relaxation in glass.
Simultaneously, depolymerized glass becomes less resistant to mechanical stress, it strength lowers \cite{Structure_strength_glasses}, and therefore the glass fracturing and crumbling accelerate.
Eventually beads of less polymerized or more stressed glass quicker deteriorate and serve for shorter time till the entire breakage.

\subsubsection{Diffusion of chemical elements} \label{subsubsec:discuss_diffusion}

The EDX analysis presented above (Section~\ref{subsec:results_EDX}) demonstrates the redistribution of alkali metal elements in glass as a result of glass deterioration.
Na and K concentrations are seen to drastically decrease in domains of degraded glass whilst they remain high in domains of undegraded glass (Figs.~\ref{fig:Beads_maps}, \ref{fig:Beads_K-Na_profiles}\,a,\,c and \ref{fig:Artificial_aging_profiles}\,b).
Moreover, line profiles of elemental distribution (Fig.~\ref{fig:Beads_K-Na_profiles}\,b,\,d) show that Na and K move from the degraded domains to undegraded ones.
Gaussian decay of K and Na concentrations in the core direction with the profiles maxima at the core-shell interface and abrupt drop of the concentrations in the shell direction point that Na and K accumulated at the interface when migrated from the shell domain, and the further diffusion of these elements took place from the interface in the direction to the bead depth and could not occur in the opposite one.
Remark, that a fissure is often detected close to the core-shell interface 
(Online Resource~\ref{esm:fig_KPbSbO5}, Fig.~ESM~\ref{esm:fig_KPbSbO5}.1\,a)
that has arisen due to stress, which had been maximum at that place.

The phenomenon of the directed migration of alkali atoms obviously cannot be an ordinary temperature-driven diffusion process.
It should be governed by some anisotropic force, e.g., by the internal stress.
We explain this observation within the above thermal-fluctuation model \cite{2020_Glas_depolimer}.

As discussed previously \cite{KSS_Electron_microscopy}, the domain around the bead orifice softened during tumbling at elevated temperature and than gradually vitrified again with bead cooling down to form the shell.
Its vitrification rate was the lowest, hence, its specific volume was also lower than that of the core after bead cooling \cite{Glass_Technology}.
This gave rise to the internal stress in the shell domain, which reached maximum at the core-shell interface. 
Thus, alkali elements diffused in the direction of the internal stress gradient (along the steepest ascent of stress) and gathered at the layer of the maximum stress.%
\footnote{%
	The strain-enhanced diffusion phenomenon is analogous to the diffusion of metals atoms that occurs during the internal gettering process often employed for the integrated circuit manufacturing  \cite{Internal_gettering-1,Internal_gettering-Multy-Si,Internal_gettering-Si-PV}
} 
In analogy with the glass depolymerization, an elementary act of the atom migration also comprise breaking of interatomic bonds complemented, however, with hops of diffusing atoms.
Bond breaking probability increases in the stress gradient direction, thus the probability of an atom hop along the steepest ascent of stress is some higher than in the opposite direction.
Hence, atoms hop some more frequently in this direction and in average a directed diffusion should occur along the steepest ascent of the internal stress, with atoms being gathered at the layer of maximum stress.  

Notice that the migration of alkali metal atoms in glass is often considered as the cause of its gradual depolymerization through bridging oxygen bond breaking and re-bonding with alkali metal cations (Na$^+$ or K$^+$) to form quadrupoles from coupled $-$O$^-$Na$^+$ ($-$O$^-$K$^+$) dipoles \cite{Nemilov_S}:
		$^{-\,\mathrm{O^-Na^+}}_{~~\,\mathrm{Na^+O^-}-}$ or $^{-\,\mathrm{O^-K^+}}_{~~\,\mathrm{K^+O^-}-}$.
%
%
%
This seems to be true for unstressed glass, yet we cannot entirely agree with this in the considered case since we believe that the alkali-metal cation diffusion and glass depolymerization processes should occur in parallel mutually intensifying each other in stressed glass.

Glass leaching undoubtedly accelerates the simplification of its structure and decreases its connectedness due to escaping cations that bind silicate units via the quadrupoles.
Simultaneously, cation escape stimulates the formation of micro discontinuities and their transformation into micro cracks owing to ion bonds breaking in the quadrupoles under the mechanical load and decay of the cation dipoles.
As a result it intensifies the glass decay process on the whole.%
\footnote{%
	It should be noted also that the binding energy of $-$O$^-$K$^+$-based quadrupoles is less than that of the $-$O$^-$Na$^+$-based ones \cite{Nemilov_S}.
	Therefore, the more K and less Na bead glass contains the more it is prone to the long-term destruction under the internal stress.
}

Other impurities, such as Cu or Pb, are not as rapidly diffusing ones as Na and K;
that is likely why they are virtually not subjected to the strain-enhanced diffusion at room temperature. 


\subsubsection{Glass density} \label{subsubsec:discuss_density}

SEM BSE images demonstrate the difference in the glass density in damaged domains of the samples and undamaged ones.
The images of undamaged or slightly damaged glass domains look dove grey, whereas those of heavily damaged regions look some darker
(Figs.~\ref{fig:Beads_samples}\,c,\,f,\,l,\,o, \ref{fig:Beads_maps_interface}\,a, \ref{fig:Sample_degradation}\,a, \ref{fig:Artificial_aging}\,c,\,d,\,f, \ref{fig:Artificial_aging_EDX}\,a and \ref{fig:Artificial_aging_profiles}).
This is because the density of glass is lower in the heavily damaged regions that is detected with SEM operated in the BSE mode.
The reason of the lower density of damaged glass lies in the outmigration of K and Na atoms from degraded domains to less deteriorated ones. 

However, despite the degraded glass is physically lighter the density of its silicate matrix is higher.
It is seen from the maps presented in Figs.~\ref{fig:Beads_maps}\,d,\,e, \ref{fig:Beads_Si-O_profiles}\,a,\,c and \ref{fig:Sample_degradation}\,g,\,h, which demonstrate an elevated number density of Si and O atoms in the degraded glass domains.

The line-scan profiles of oxygen and silicon concentrations obtained in the vicinity of the core-shell interface near the orifice of the sample 1 show the content of these elements to be constant in the core (Fig.~\ref{fig:Beads_Si-O_profiles}\,b,\,d).
An approximately 10~{\textmu}m thick layer of rarefied glass is detected close to the interface.
A considerable tensile strain induced by the stress that reaches maximum at the core-shell interface is likely responsible for the appearance of this layer of stretched glass. 
The position of this layer coincides with a typical distance from a bead orifice, at which a circular fissure around the hole is often observed in strongly cracked beads 
(Online Resource~\ref{esm:fig_KPbSbO5}, Fig.~ESM~\ref{esm:fig_KPbSbO5}.1\,a).

In the shell domain, the concentrations of O and Si atoms rapidly grow when moving away from the core-shell interface and become constant again soon, yet at a much higher level than in the core domain (Fig.~\ref{fig:Beads_Si-O_profiles}\,b,\,d).
Thus, glass silicate matrix is remarkably denser in the damaged shell layer around the orifice.
This well agrees with our assumption that glass re-vitrified slowly after tumbling at elevated temperature in this domain and then cooled down slower than glass did in the core region.

\subsubsection{Glass discolouration} \label{subsubsec:discuss_discolor}
	
The changes in the glass colour (Figs.~\ref{fig:Beads_stress}\,b,\,c and \ref{fig:Beads_samples}), i.e. its greening and yellowing \cite{Yuryev_JOPT}, might be tentatively attributed to the migration of alkali metals, K and Na, from some domains and the formation of regions depleted in these elements. 
Since the Cu concentration remains unchanged in these regions, the reduction of the alkali metals concentration likely might affect the glass colour decreasing the connectedness of glass units due to escaping cations that bind silicate units via the quadrupoles.
This might change the coordination of Cu$^{2+}$ ions in such a way changing its absorption band and hence the glass colour.
Alternatively, the depolymerization of glass in the degraded domains may change the Cu$^{2+}$ coordination.

Notice also, that peaks assigned to Cu$_2$O (Cu$^+$) were previously detected in photoluminescence (PL) and Raman spectra recorded in domains of changed (yellow-green or brownish) colour, with the Cu$_2$O PL intensity increased with the change of the glass structure from Q$^3$ to Q$^2$ \cite{Beads_degraded_areas-2019}.
Besides, PL assigned to O$^+$ and O$^{2+}$ vacancies and a yellow PL peak of exciton were also observed.
This seems to be a highly probable cause of the observed glass discolouration. 

The third possible reason of the glass colour changes is the growth of the Rayleigh scattering intensity due to the glass degradation, which we reported in Ref.~\cite{Yuryev_JOPT}, that was likely caused by the increasing number density of micro discontinuities and micro cracks (as well as the sizes of the latter) \cite{KSS_Electron_microscopy}, which are effective light scatterers 
\cite{Yuryev_JOPT,2020_Glas_depolimer}.%
\footnote{%
Note that the characteristic dimensions $a$ of the light scatterers (they were estimated as $a \ll 150$~nm in Ref.~\cite{Yuryev_JOPT}), the number density of which grows with the rising degree of glass corrosion, coincide with the sizes of nucleating micro discontinuities that, according to Ref.~\cite{Microcracks_under_loading}, are typical defects emerging due to the thermal fluctuation process at locations of shear and rotational strains. 
Previously, we have also reported on the detection with a transmission electron microscope of some tiny clusters, 1 to 4~nm in size, in this kind of glass \cite{KSS_Electron_microscopy}, which may either arise due to the effect of an e-beam or be inherent to this glass and visualized under the intense e-beam irradiation. 
}
%
The light transmission and scattering spectra presented in Ref.~\cite{Yuryev_JOPT} evidence for the third mechanism of the glass discolouration.

Darkening of the heavily degraded glass domains \cite{Yuryev_JOPT} is assigned to the high number of fractures, as it was demonstrated by our experiments on immersion of a blackened bead in highly refined liquid paraffin \cite{Beads_KSS_SPIE}.%
\footnote{%
	After the immersion in mineral oil and treating with ultrasound (120~W, 40~kHz) for some time, the cracked surface region of the darkened seed bead considerably brightened \cite{Beads_KSS_SPIE}.%
}


\subsection{Gradual destruction of glass after polishing}  \label{subsec:discuss_polishing}

The above model might also explain the phenomenon of the accelerated destruction of glass after the thin section preparation (Section~\ref{subsec:sample_storage}).
It is commonly known that, as a result of sample grinding and polishing during the treatment, numerous extended micro defects, such as, e.g., micro scratches, micro cracks, micro voids, digs or pits, {\it etc.} \cite{Polishing_micro-defects_glass}, are generated in the near-surface layers that form a damaged layer \cite{Semicond_Meas_Instrum}, which can be quite thick (it penetrates to the depth of tens micrometers in some materials \cite{depth_mechanical_damage}).
Besides, additional non-uniformly distributed internal stress is introduced by this processing into the near-surface layer \cite{Semicond_Meas_Instrum}.
Both these factors---defects and stress---may cause the accelerated destruction of glass in this layer.
The general picture of this phenomenon may be as follows.

The introduced extended defects give rise to accelerated glass cracking under the internal stress.
Micro cracks, voids and discontinuities grow forming new cracks and relieving the stress starting from the surface, where the defect density is maximum, and gradually spreading the stress relaxation into the bulk through the formation a degraded glass layer.
Atoms of alkali metals move along stress gradient deeper into glass, since atom hops in this direction is a little more probable than in the opposite one due to lowering breaking barrier of atomic bonds \cite{Atomic-Level_Fracture_Solids} in the stress gradient direction, and gradually escape from the surface (Fig.~\ref{fig:Sample_degradation}).
Simultaneously, glass depolymerizes faster due to excess stress (Fig.~\ref{fig:Sample_degradation_FTIR}) and the out-migration of alkali metal atoms; this also relieves the stress but increases the silicate matrix density (Fig.~\ref{fig:Sample_degradation}).
In such a way, a front of glass destruction gradually moves from the surface into the depth causing stress relief through the glass depolymerization and cracking accompanied by the directed migration of alkali metal atoms into the depth of a bead; the degraded glass layer broadens until reaches the thickness of the polishing-damaged near-surface layer.

\subsection{Artificial ageing} \label{subsec:discuss_aging}

Experiment on the annealing of intact and heavily degraded beads at the temperature of {300\textcelsius} for 15~minutes in the atmospheric air (Section~\ref{subsec:results_aging}) has given the following results.

First, a shell composed of scales separated with cracks has arisen on a previously intact bead (Fig.~\ref{fig:Artificial_aging}).
As a result, the bead has become resembling the heavily degraded one.
At the same time, the latter one has not changed visually after the anneal, yet it has become more fragile and started rapid crumbling.

Second, Na and K has moved out of the newly formed shell of the previously intact bead (Table~\ref{tab:EDS-2} and Fig.~\ref{fig:Artificial_aging_EDX}\,b,\,c) and glass of its core has become heavier then that of the shell (Figs.~\ref{fig:Artificial_aging}\,c,\,d, ~\ref{fig:Artificial_aging_EDX}\,a and ~\ref{fig:Artificial_aging_profiles}).
However, in contrast to the long-term corrosion at room temperature, we did not observe any changes in Si and O content in this sample after the heat treatment (Figs.~\ref{fig:Artificial_aging_EDX}\,d,\,e and \ref{fig:Artificial_aging_profiles}\,b).
(Some difference seen in Table~\ref{tab:EDS-2} may be because of the spatial non-uniformity of glass properties within the frame rather than due to glass alteration caused by the annealing.)
Hence, the silicate matrix density has remained unchanged in spite of the anneal.
Besides, unlike the case of the long-term corrosion, Pb atoms are seen to have migrated from the shell due to the thermal processing (Fig.~\ref{fig:Artificial_aging_EDX}): their content in the core is seen to be higher than in the shell.

Finally, glass has depolymerized in the shell of the initially undamaged bead in the process of the heat treatment (Fig.~\ref{fig:Artificial_aging_FTIR}\,a,\,b) and remained unchanged in the initially depolymerized heavily degraded one (Fig.~\ref{fig:Artificial_aging_FTIR}\,b,\,c).
In the former case, the strong bands peaked at 930 to 950~cm$^{-1}$ and around 1080~cm$^{-1}$ assigned to vibrations of rings and/or chains (Q$^2$) and sheets (Q$^3$), respectively, as well as somewhat weaker bands peaked close to 1010~cm$^{-1}$, 1140~cm$^{-1}$ and at 1150 to 1200~cm$^{-1}$ assigned to vibrations of chains (Q$^2$), sheets (Q$^3$) and frameworks (Q$^4$), respectively, transform to the only very intense peak at $\sim$1000~cm$^{-1}$ assigned to vibrations of chains \cite{glass_units,MOCIOIU-2013}.
In the case of the strongly degraded bead, bands associated with vibrations of chains absolutely predominate in the spectra before and after the annealing.

Thus, bead glass can be artificially aged by a heat treatment for a quite short time at a moderate temperature.
This may be accounted for using the above model of the stress-driven corrosion.
A bead is coated with a layer of a modified glass, which we refer to as shell, that is somewhat quenched at the outer area during cooling after the tumbling at an elevated temperature.
This glass is compressively stained \cite{GLASS_AN_USSR}.
Bead rapid heating and cooling results in the development of additional stresses in this layer during the thermal cycling (likely quite strong ones) that causes the fast outmigration of some impurity atoms, especially atoms of alkali metals and, as one can see, lead, and dramatically accelerates glass depolymerization and fracturing.
In this case, we also consider the thermal-fluctuation mechanism discussed above as a driving force of the glass alteration during the annealing.

\subsection{General conclusions} \label{subsec:concl}

In conclusion, we would like to emphasize that the most important role in the corrosion of bead glass is played by internal stresses that have arisen in the glass during the manufacture of beads since the corrosion of the glass beads takes place owing to gradual relaxation of that stress.  

The overall picture of the destruction of glass beads of the 19th century looks as follows \cite{RSF_16-18-10366/2018}.

 


\textit{The nucleation of crystallites in turquoise beads.} 
When glass was melted at $T>1200$\,{\textcelsius}, orthorhombic KSbOSiO$_4$ crystallites were nucleating in the melt \cite{Yur_JAP,Beads_KSS_SPIE,KSS_Electron_microscopy,GosNIIR_Conference}.

\textit{The growth of crystallites and the softening of the near-surface region of glass.} 
In the process of tumbling of chopped pieces of a glass tube at an elevated temperature, crystals of orthorhombic KSbOSiO$_4$ grew in glass and the surface layer of glass softened \cite{KSS_Electron_microscopy}. 
Due to the difference in the efficiency of heat exchange between the surface and the environment in different parts of the beads---in outer parts of the beads, heat transfer was more efficient than inside a hole filled with a grinding paste of mixed clay, charcoal and chalk \cite{Yurova_large}---the softened layer was significantly thicker inside the hole and in close proximity to it than in the outer part of the beads \cite{KSS_Electron_microscopy}.

\textit{The occurrence of mechanical stresses in glass and the formation of microcracks.} 
When the beads cooled down after tumbling, due to the difference in the linear expansion coefficients of orthorhombic KSbOSiO$_4$ and glass, tensile stresses were arising in the latter, and micro discontinuities and microcracks were appearing in the core region of glass; in addition, the near-surface layer was re-vitrifying \cite{KSS_Electron_microscopy}. 

The silicate composition and density of the modified glass are determined by its cooling rate during the vitrification \cite{Glass_Technology}. 
Due to the difference in the silicate composition of glass of the inner (core) region and the modified shell layer, glass in these regions had different linear expansion coefficients, as a result of which a mechanical stress occurred in glass near the interface of these regions, which extended both to the inner region and to the entire layer of modified glass; microcracks nucleated in tense glass. 
Since the layer of modified glass shell inside and near the orifice was much thicker than in the outer region, and the cooling rate of the outer part of the modified layer, due to more efficient heat exchange, was much higher than the cooling rate of the modified layer inside and near the orifice, a higher internal stress should accumulate in heavier glass of the latter during the cooling process than in the outer part of the shell layer.%
\footnote{%
Note also that, since the cooling rate of the outer shell was greater than that of the core and the cooling rate of the shell around the orifice was less then that of the core, glass has become compressively stressed in the outer shell, as if a bead had been quenched, whereas glass has become stretched in the shell around the orifice \cite{GLASS_AN_USSR}.
Sometimes, the tension eventually gave rise to the formation of a crack at the core-shell interface that ran around the bead hole at some distance from it.
}
That is why the microcracks nucleation process should go faster in the layer of the modified glass shell inside and near the hole than in the outer shell region of beads.

\textit{The spread of cracks and the formation of grains.} 
After beads were cooled down, all relatively fast processes in them stopped; however, considerable internal stresses remained in their glass. 
As a result, under the effect of internal mechanical stress, a slow process of the nucleation and the formation of microcracks continued in it even at room temperature; the cracks grew, connected in networks, increased in size, and over time, formed isolated non-stressed areas, or, in other words, gave rise to glass grains resembling sand particles \cite{KSS_Electron_microscopy}. 
The region of modified glass shell was often completely (near the orifice) or partially separated by a crack from the core region. 

The rate of the processes of nucleation and growth of cracks, and therefore, the magnitude of the internal stress that generates them, determines the rate of destruction of the beads, i.e. their lifetime till the complete decay.

\textit{The depolymerization of glass.} 
Simultaneously with the nucleation and growth of cracks under internal stress in glass of beads, the depolymerization process, i.e. the simplification of the silicate composition, proceeded. 
Due to the depolymerization of glass, its resistance to mechanical stress decreased, the strength reduced \cite{Structure_strength_glasses}, and therefore the formation and growth of cracks accelerated.

\textit{Glass leaching and discolouration.} \label{par:discoloration}
As an explanation of the gradual change in the colour of glass in the process of its corrosion, as well as the reduced concentration of Na and K in strongly damaged regions, it is assumed by analogy with crystals that the mechanism of the accelerated diffusion of impurities under the effect of the internal stress is valid for glass.
Since the originally most stressed areas of glass have destroyed to the greatest extent, it is reasonable to assume that even long before the complete stress relaxation through the formation of cracks, the redistribution of impurities had occurred in them: alkali metals had been gradually displaced from them. 
The alteration of the glass colour in them to green or yellow may be due to the migration of K and Na or due to changes of the silicate composition and, as a consequence, the change in the coordination of copper atoms. 
Cu$_2$O forming with the glass degradation may also cause appearing green, yellow or brown tint of glass \cite{Beads_degraded_areas-2019}.

Another reason of the changes in glass colour, which seems to us to be the most probable, is the increase in the Rayleigh scattering intensity in course of the glass degradation \cite{Yuryev_JOPT}, that is likely caused by the increasing number density of micro discontinuities and micro cracks (and the dimensions of the latter) \cite{KSS_Electron_microscopy}, which effectually scatter light \cite{Yuryev_JOPT,2020_Glas_depolimer}.%

\textit{The crack nucleation and growth mechanism. The depolymerization mechanism.} \label{par:depolymerization}
Both these processes are caused by common elementary processes of breaking loaded interatomic bonds by local fluctuations of the vibrational energy (the so-called thermal-fluctuation mechanism) \cite{2020_Glas_depolimer,Atomic-Level_Fracture_Solids}. 
When an interatomic bond is loaded, the energy barrier for its rupture decreases significantly, as a result of which the probability that the local fluctuation of the vibrational energy of this bond arising at some moment in time exceeds this barrier increases; accordingly, the likelihood of the bond breaking also increases. 
The more a bond is loaded, the lower is the barrier for its breaking and, consequently, the greater is the frequency of appearance and hence the average density of energy fluctuations of atomic bond oscillations, which exceed the magnitude of this barrier, in the bulk of glass. 
Thus, the stronger the interatomic bonds are loaded (the higher the internal stress is), the more often they break, and the faster the process of formation of microcrack nuclei or the process of appearance and growth of microcracks as well as that of the formation of extended cracks.  

Elementary acts of the glass depolymerization process are also the breaking of interatomic bonds, i.e. all that has been said above about the nucleation of cracks also fully applies to the glass depolymerization process: this is also explained from the standpoint of the kinetic theory of solids within the framework of the thermal-fluctuation mechanism \cite{2020_Glas_depolimer}. 
The depolymerization of stressed glass and the formation of cracks in it lead to a gradual relaxation of the internal stress in it through its slow destruction.  
The rate of both these processes are strongly dependent on temperature.
However, even at the temperature, at which beaded artefacts usually existed in everyday life or articles made of beads were kept at a museum, i.e. at room temperature or close to it, these processes were fast enough to lead to the almost entire destruction of turquoise beads for one or two centuries.

\textit{The stress-driven diffusion mechanism.} \label{par:diffusion}
Alkali metal atoms, Na and K, gradually migrate from some domains to others even if the latter are situated in the core of a seed bead.
This phenomenon, as mentioned above, cannot be a routine temperature-driven diffusion process but should be mainly governed by a directional driving force, i.e. by the internal stress.

This phenomenon is also explained within the framework of the thermal-fluctuation mechanism model \cite{2020_Glas_depolimer}.
Like it has been suggested for the explanation of glass cracking and depolymerization, elementary acts of this process also comprise breaking of interatomic bonds yet complemented by hops of migrating atoms.
Bond breaking likelihood grows in the stress gradient direction, thus the probability of an atom hop in this direction a little higher than in the opposite one.
Consequently, atoms should move some more frequently in this direction and in average a directed diffusion should take place along the steepest ascent of the internal stress, with atoms being accumulated at the domain of its maximum magnitude. 

Glass leaching accelerates the simplification of its structure and decreases its connectedness since cations that bind silicate units escape.
Alkali metal cation escape also promotes the formation of micro discontinuities and their transition into cracks since ion bonds break in the quadrupoles due to mechanical load and oxygen-cation dipoles decay.
As a result glass leaching facilitates the process of its decay.

\section{Summary} \label{sec:Summary}

{{In summary}}, we should conclude that the phenomenon of the long-term corrosion of 19th century beads made of turquoise lead-potassium glass comprises three parallel but mutually intensifying processes, which are inherent to this phenomenon and are controlled by the common driving force, namely, by the internal stress originated from the bead production process.
These three processes are the directed impurity migration, the glass depolymerization, and the nucleation of micro discontinuities.
In course of time, their cumulative and apparently synergistic effect results in glass cracking terminated with bead crumbling into tiny discoloured sand-like particles.
The thermal-fluctuation mechanism accounts for this phenomenon at the atomic level.  

We should emphasize that the approach based on the thermal fluctuation mechanism can be applied not only for the description of the long-term corrosion of stressed tiny glass seed beads.
It is much more widely applicable.
This approach can be used to describe degradation and destruction processes in various solid substances and composite materials under internal or external stresses such as technical or artistic glasses, glass fibres and fibreglass composites, ceramics and glazes, polymers, and even crystalline materials such as, e.g., metals or semiconductor structures.

It may also be important for predicting long-term stability of technical glasses as well as for synthesis of crystallite/glass composites.
Besides, it can be useful for studying the corrosion resistance and durability of vitreous and polymer materials used to immobilize or encapsulate hazardous wastes.%
\footnote{%
	Notice, that earlier in Ref.~\cite{stained_glass-waste_matrices}, historical and archaeological stained glasses were used to estimate the effect of weathering conditions and glass composition on glass dissolution on a millennium time scale that enabled the assessment of the possible long-term environmental and compositional effect on the behaviour of vitrified waste matrices, which could be used for careful developing a glass formulation strategy for vitrified nuclear wastes.     
	The current study also paves a novel way to the evaluation of the long-term stability of vitrified wastes, but as distinct from the former ones taking into account the long-term effect of internal stresses on glass structure causing its decay.  %
}

The stress driven diffusion in solids, e.g., the diffusion and intermixing in nanostructures, such as heterostructures with quantum dots or quantum wells, can also be explained in terms of the thermal fluctuation mechanism \cite{GE_QD-strain_above}.



{\newpage}

\clearpage

\section*{CRediT authorship contribution statement} 

\textbf{Irina F. Kadikova:}
Investigation,
Formal analysis, 
Writing\,-- Review
\& Editing.
\textbf{Tatyana V. Yuryeva:}
Conceptualization,
Funding acquisition,
Project administration,
Resources,
Data Curation,
Investigation,
Visualization.
\textbf{Ekaterina A. Morozova:}
Investigation,
Writing\,-- Review
\& Editing.
\textbf{Irina A. Grigorieva:}
Investigation,
Writing\,-- Review
\& Editing.
\textbf{Ilya B. Afanasyev:}
Investigation.
\textbf{Vladimir Y. Karpenko:}
Investigation.
\textbf{Vladimir A. Yuryev:}
Conceptualization,
Methodology,
Funding acquisition, 
Investigation,
Formal analysis,
Writing\,--
Original draft, 
Writing\,-- Review
\& Editing,
Supervision.



\section*{Acknowledgments}

The Russian Science Foundation funded this research via the grant number 16-18-10366. 
The research was accomplished under the non-profit collaboration agreement between GOSNIIR and GPI RAS.

\small{
	\paragraph{\small\bf Conflicts of interest} 
	No conflicts of interest exist,  which could potentially influence the work.
}
\normalsize 

{\newpage}

\clearpage
 
 
 \section*{Electronic supplementary materials} \label{sec:ESM}
 
\newcounter{global}
\newcounter{esm} 
\newcounter{ESM_fig_number}[global]
\newcounter{movie}[global]
\newcounter{fig_FTIR_Epoxy}[global]
\newcounter{fig_FTIR_All_Spectra}[global]
\newcounter{fig_KPbSbO5}[global]
\newcounter{ESM_tab_number}[global]

\paragraph*{Online Resource \addtocounter{esm}{1}\arabic{esm}: video file Electronic\_supplementary\_material\_\arabic{esm}.mp4  \label{par:esm_\arabic{esm}}}~
\refstepcounter{global}
\setcounter{movie}{\value{esm}}
\label{esm:movie}
{%
\\
	Degraded translucent turquoise glass seed beads instantly break, with glass pieces flying away, when touched with a needle whereas similar intact ones, adjacent in a beadwork, do not even if intensely touched with a needle  \cite{2022_Glas_alter_ESM-1}. 
	This is due to high internal stress accumulated in the deteriorated beads during making as well as low strength of the fractured glass. 
	The intact beads are not as stressed and do not contain cracks, thus they are much stronger than the deteriorated ones.
}%

\paragraph*{Online Resource \addtocounter{esm}{1}\arabic{esm}: file Electronic\_supplementary\_material\_\arabic{esm}.pdf \label{par:esm_\arabic{esm}}}~
\refstepcounter{global}
\setcounter{ESM_fig_number}{0}
\setcounter{ESM_tab_number}{0}
\setcounter{fig_KPbSbO5}{\value{esm}}
\label{esm:fig_KPbSbO5}
{
	\\		Appendix~A. IR absorption bands of orthorhombic KSbOSiO$_4$ crystals \cite{KSbOSiO4_IR_absorption}.
}%
{
	\\ \addtocounter{ESM_fig_number}{1}
	Figure~ESM~\arabic{esm}.\arabic{ESM_fig_number}: 
	SEM SE and BSE images of unstable turquoise glass beads of the 19th century fragmented due to internal corrosion (a,\,b)  
	and 
	KSS crystals in glass of those beads (c--g); 
	figures~1 indicate KSS colonies in glass,
	numerals~2 and 3 show core and shell domains of glass, respectively;
	SE images show mainly the spatial relief, BSE images represent mainly the substance density. 
	The KSS colonies are seen to cause glass fracture because of mechanical stresses built in glass during its cooling after tumble finishing of the beads as discussed in Ref.~\cite{KSS_Electron_microscopy}.
}%
\newcounter{Fig \arabic{ESM_fig_number}}
\setcounter{Fig \arabic{ESM_fig_number}}{\value{ESM_fig_number}}
{
	\\ \addtocounter{ESM_fig_number}{1}
	Figure~ESM~\arabic{esm}.\arabic{ESM_fig_number}:
	SEM SE image of the synthesized orthorhombic KSbOSiO$_4$ powder that was used as a standard in the experiments of this work. 
}%
\newcounter{Fig \arabic{ESM_fig_number}}
\setcounter{Fig \arabic{ESM_fig_number}}{\value{ESM_fig_number}}
{
	\\ \addtocounter{ESM_fig_number}{1}
	Figure~ESM~\arabic{esm}.\arabic{ESM_fig_number}: 
	FTIR absorption spectra of orthorhombic KSbOSiO$_4$:
	a survey spectrum (a);
	a mid-IR spectrum (b), characteristic spectral bands are marked with numerals from 1 to 8, letter M in band designation means that it belongs to the mid-infrared spectral range;
	a far-IR spectrum (c), characteristic spectral bands are marked with numerals from 1 to 9, letter F in band designation means that it belongs to the far-infrared spectral range.
}%
\newcounter{Fig \arabic{ESM_fig_number}}
\setcounter{Fig \arabic{ESM_fig_number}}{\value{ESM_fig_number}}
{
	\\ \addtocounter{ESM_fig_number}{1}
	Figure~ESM~\arabic{esm}.\arabic{ESM_fig_number}:
	Peak analysis of mid-IR absorption bands of orthorhombic KSbOSiO$_4$.
}%
\newcounter{Fig \arabic{ESM_fig_number}}
\setcounter{Fig \arabic{ESM_fig_number}}{\value{ESM_fig_number}}
{
	\\ \addtocounter{ESM_fig_number}{1}
	Figure~ESM~\arabic{esm}.\arabic{ESM_fig_number}:
	Peak analysis of far-IR absorption bands of orthorhombic KSbOSiO$_4$. 
}%
\newcounter{Fig \arabic{ESM_fig_number}}
\setcounter{Fig \arabic{ESM_fig_number}}{\value{ESM_fig_number}}
{
	\\ \addtocounter{ESM_tab_number}{1}
	Table~ESM~\arabic{esm}.\arabic{ESM_tab_number}:
	Spectral bands of mid-IR radiation absorption in orthorhombic KSbOSiO$_4$ crystals (Fig.~ESM~\arabic{esm}.4). 
}%
\newcounter{tab \arabic{ESM_tab_number}}
\setcounter{tab \arabic{ESM_tab_number}}{\value{ESM_tab_number}}
{
	\\ \addtocounter{ESM_tab_number}{1}
	Table~ESM~\arabic{esm}.\arabic{ESM_tab_number}:
	Spectral bands of far-IR radiation absorption in orthorhombic KSbOSiO$_4$ crystals (Fig.~ESM~\arabic{esm}.5).
}%
\newcounter{tab \arabic{ESM_tab_number}}
\setcounter{tab \arabic{ESM_tab_number}}{\value{ESM_tab_number}}

\paragraph*{Online Resource \addtocounter{esm}{1}\arabic{esm}: file Electronic\_supplementary\_material\_\arabic{esm}.pdf  \label{par:esm_\arabic{esm}}}~
\refstepcounter{global}
\setcounter{ESM_fig_number}{0}
\setcounter{fig_FTIR_All_Spectra}{\value{esm}}
\label{esm:fig_FTIR_All_Spectra}
{
	\\ \addtocounter{ESM_fig_number}{1}
	Figure~ESM~\arabic{esm}.\arabic{ESM_fig_number}: 
	Sample~\arabic{ESM_fig_number}: 
	a micro photograph, analysed areas and IR absorption spectra;
	the numbered rectangles show the areas	of FTIR spectral analysis;
	empty triangles depict the experimental data,
	olive and blue lines present peaks derived as a result of the spectra deconvolution 
	(the blue lines show the strongest absorption peaks),
	red lines are the cumulative fit curves;
	moderate and weak peaks composing the absorption bands at the wave numbers below 840~cm$^{-1}$ are omitted \cite{2022_Glas_alter_ESM-3}.
}%
															\newcounter{fig \arabic{ESM_fig_number}}
															\setcounter{fig \arabic{ESM_fig_number}}{\value{ESM_fig_number}}
{
	\\ \addtocounter{ESM_fig_number}{1}
	Figure~ESM~\arabic{esm}.\arabic{ESM_fig_number}:
		Sample~\arabic{ESM_fig_number}: 
	a micro photograph, analysed areas and IR absorption spectra; 
	the designations are the same as in Fig.~ESM~\arabic{esm}.\arabic{fig 1}. 
}%
															\newcounter{fig \arabic{ESM_fig_number}}
															\setcounter{fig \arabic{ESM_fig_number}}{\value{ESM_fig_number}}
{
	\\ \addtocounter{ESM_fig_number}{1}
	Figure~ESM~\arabic{esm}.\arabic{ESM_fig_number}:
	Sample~\arabic{ESM_fig_number}: 
	a micro photograph, analysed areas and IR absorption spectra; 
	the designations are the same as in Fig.~ESM~\arabic{esm}.\arabic{fig 1}. 
}%
															\newcounter{fig \arabic{ESM_fig_number}}
															\setcounter{fig \arabic{ESM_fig_number}}{\value{ESM_fig_number}}
{
	\\ \addtocounter{ESM_fig_number}{1}
	Figure~ESM~\arabic{esm}.\arabic{ESM_fig_number}: 
	Sample~\arabic{ESM_fig_number}: 
	a micro photograph, analysed areas and IR absorption spectra; 
	the designations are the same as in Fig.~ESM~\arabic{esm}.\arabic{fig 1};
	additionally,
	the pink line presents a spectrum from the area with the dominating absorption by orthorhombic KSbOSiO$_4$ (KSS) crystals,
	the royal blue line and the empty circles show the KSS reference sample spectrum.
}%
															\newcounter{fig \arabic{ESM_fig_number}}
															\setcounter{fig \arabic{ESM_fig_number}}{\value{ESM_fig_number}}
{
	\\ \addtocounter{ESM_fig_number}{1}
	Figure~ESM~\arabic{esm}.\arabic{ESM_fig_number}:
	Sample~\arabic{ESM_fig_number}: 
	a micro photograph, analysed areas and IR absorption spectra; 
	the designations are the same as in Figs.~ESM~\arabic{esm}.\arabic{fig 1} and~ESM~\arabic{esm}.\arabic{fig 4}. 
}%
															\newcounter{fig \arabic{ESM_fig_number}}
															\setcounter{fig \arabic{ESM_fig_number}}{\value{ESM_fig_number}}
{
	\\ \addtocounter{ESM_fig_number}{1}
	Figure~ESM~\arabic{esm}.\arabic{ESM_fig_number}:
	Sample~\arabic{ESM_fig_number}: 
	a micro photograph, analysed areas and IR absorption spectra; 
	the designations are the same as in Fig.~ESM~\arabic{esm}.\arabic{fig 1}. 
}%
															\newcounter{fig \arabic{ESM_fig_number}}
															\setcounter{fig \arabic{ESM_fig_number}}{\value{ESM_fig_number}}
{
	\\ \addtocounter{ESM_fig_number}{1}
	Figure~ESM~\arabic{esm}.\arabic{ESM_fig_number}:  
	Sample~\arabic{fig 4}: 
	IR absorption spectra recorded in a few years after the sample preparation 
	(compare with Fig.~ESM~\arabic{esm}.\arabic{fig 4});
	the designations are the same as in Fig.~ESM~\arabic{esm}.\arabic{fig 1};
	additionally,
	the royal blue line and the empty circles show the KSS reference sample spectrum;
	the sample micro photograph and the analysed areas are shown in Fig.~ESM~\arabic{esm}.\arabic{fig 4}.
}%
															\newcounter{fig \arabic{ESM_fig_number}}
															\setcounter{fig \arabic{ESM_fig_number}}{\value{ESM_fig_number}}
															


\clearpage

{\newpage}

\clearpage

\section*{Data Availability} 
The raw/processed data required to reproduce these findings cannot be shared at this time as the data also forms part of an ongoing study.


{\newpage}

\clearpage

\begin{table}[t]
	\caption{%
		Relative content of some chemical elements in the samples at different sites on the sections (EDX analysis).
		Site (domain) preservation degree is graded into three levels: $u$ is an {\it undamaged} or slightly damaged blue site, $m$ is a  {\it moderately} damaged or yellow site and $h$ is a {\it heavily} damaged or dark site.	
	} 
	\label{tab:EDS-1}
	\scriptsize 
	\begin{minipage}{\textwidth}{\hspace{0cm}}
		\begin{tabular}{cccccccccccccc} 
			\noalign{\smallskip}\hline
			{Chemical}	& \multicolumn{13}{c}{Sample number (\textit{site preservation degree}) } \\ \cline{2-14}
			element		& 1 ($u$)&2 ($u$) &2 ($m$) &2 ($h$)&3 ($u$)&3 ($h$)&4 ($m$)&4 ($h$)&5 ($m$)&5 ($h$) &6 ($u$)&6 ($m$)&6 ($h$)\\ \cline{2-14}
			& \multicolumn{13}{c}{Relative content (at.\%)} \\ \hline
			Na				& 8.44  	& 7.41 	& 2.62   & 2.23  & 7.39  & 2.40  & 3.04  & 1.85  & 3.74  & 3.35   & 6.44  & 1.74  &		   \\
			Si				& 53.53 	& 57.63  & 73.95 	& 71.70 & 56.97 & 74.36 & 73.18 & 77.76 & 72.62 & 77.69  & 59.25 & 73.26 & 54.27 \\
			K				& 27.60 	& 27.18 	& 13.08	& 13.09 & 26.88 & 13.12 & 19.47 & 14.85 & 16.13 & 12.90  & 28.66 & 15.94 & 24.15 \\
			Ca				& 2.53  	& 1.09   & 1.83	& 3.17  & 1.51  &	1.66	&  	  & 1.06	 & 1.18  & 1.18   & 		  & 0.66  &		   \\	
			Cu				& 2.81  	& 2.13   & 1.93	& 2.35  & 1.98	 & 2.12	& 2.50  & 2.50  & 1.22  & 1.45   & 3.82  & 4.45  & 3.52  \\
			Sb          & 1.82   & 1.31   & 2.54   & 3.64  & 2.12  & 2.15  & 0.98  & 1.05  & 3.59  & 1.99   &       & 1.30  & 18.06 \\
			Pb				& 3.27  	& 3.25   & 4.05 	& 3.82  & 3.15  & 4.19	& 0.83  & 0.93  & 1.52  & 1.44   & 1.83  & 2.65  &	      \\
			\hline  
		\end{tabular} 
	\end{minipage}
\end{table}

\begin{table}[t]
	\begin{minipage}[l]{1\textwidth}{\vspace{0cm}}{\hspace{0cm}}
		\caption{%
			Relative content of some chemical elements at different sites on the annealed sample (Fig.~\ref{fig:Artificial_aging_EDX}).
			The  data of EDX analysis are presented for an area on the shell in the centre of Fig.~\ref{fig:Artificial_aging_EDX}\,a and two areas on the core (in the lower left and lower right windows (LLW and LRW), Fig.~\ref{fig:Artificial_aging_profiles}\,a).
		} 
		\label{tab:EDS-2}
			\begin{tabular}{cccc} 
				\noalign{\smallskip}\hline
				{Chemical }	& \multicolumn{3}{c}{Relative content (at.\%)} \\ \cline{2-4}
				element & shell & \multicolumn{2}{c}{core}  \\ \cline{3-4}
				& & LLW area& LRW area\\ \hline
				Na				&2.07		&7.11		&7.60 	\\
				Si				& 75.34	&63.68	&61.39 	\\
				K				& 10.93	&20.17	&20.71 	\\
				Ca				& 4.92	&1.71		&2.56		\\	
				Cu				& 2.76	&2.81		&2.92		\\
				Sb          & 0.59	&			&0.55  	\\
				Pb				& 3.39	&4.52		&4.27 	\\
				\hline  
			\end{tabular} 
		\end{minipage}
	\end{table}
	

\label{END_of_MANUSCRIPT}

\end{document}